\documentclass[aps,prd,longbibliography,preprint,amsmath,amssymb,floatfix]{revtex4-1}
\usepackage{graphicx}

\usepackage{bm}% bold math
\usepackage{xcolor}
\usepackage{url}
\usepackage{bookmark}
\usepackage{lineno}

\newcommand{\be}{\begin{equation}}
\newcommand{\ee}{\end{equation}}
\newcommand{\eq}[1]{Eq.~(\ref{#1})}
\newcommand{\fig}[1]{Fig.~\ref{#1}}
\def\bea{\begin{eqnarray}}
\def\eea{\end{eqnarray}}

\def\vq{{\bf q}}

\def\vk{{\bf k}}

\def\vr{{\bf r}}

\begin{document}

\title{Theory of charge dynamics in bilayer electron system with long-range Coulomb interaction} 

\author{Hiroyuki Yamase}
\affiliation{
Research Center of Materials Nanoarchitectonics (MANA), National Institute for Materials Science (NIMS), Tsukuba 305-0047, Japan}

\date{November 13, 2024}

\begin{abstract} 
We perform a comprehensive study of charge excitations in a bilayer electron system in the presence of the long-range Coulomb interaction (LRC). Our major point is to derive formulae of the LRC that fully respect the bilayer lattice structure.  This is an extension of the LRC obtained by Fetter in the electron-gas model 50 years ago and can now be applicable to any electron density. We then provide general formulae of the charge susceptibility in the random phase approximation and study them numerically. The charge ordering tendency is not found and instead we find two plasmon modes, $\omega_{+}$ and $\omega_{-}$ modes. Our second major point is to  elucidate their spectral weight distribution and the effect of electron tunneling between the layers. The spectral weight of the $\omega_{\pm}$ modes does not have $2\pi$ periodicity along the $q_{z}c$ direction; $q_z$ and $c$ are momentum and the lattice constant along the $z$ direction, respectively. The $\omega_{+}$ mode loses spectral weight at in-plane momentum $\vq_{\parallel}=(0,0)$ at $q_{z}c=2n\pi$ with $n$ being integer whereas the $\omega_{-}$ mode has no spectral weight at $q_{z}c=0$ for any $\vq_{\parallel}$ but acquires sizable spectral weight at  $q_{z}c=2n\pi$ with $n \ne 0$. Both $\omega_{\pm}$ modes are gapped at $\vq_{\parallel}=(0,0)$. When $q_{z}c$ is away from $2n\pi$, the $\omega_{\pm}$ modes show striking behavior. When the intrabilayer hopping $t_z$ is relatively small (large), the $\omega_-$ ($\omega_+$) mode becomes gapless at $\vq_{\parallel}=(0,0)$ whereas the $\omega_+$ ($\omega_-$) mode retains the gap. However, when the interbilayer hopping integral $t_{z}'$ is taken into account, the gapless mode acquires a gap at $\vq_{\parallel}=(0,0)$ and both $\omega_{\pm}$ modes are gapped at any $q_{z}c$. To highlight the special feature of the LRC, we also clarify a difference to the case of a short-range interaction. While the strong electron correlation effects are not included, the present theory captures available data of the charge excitations observed by resonant inelastic x-ray scattering for Y-based cuprate superconductors. \end{abstract}

%\pacs{74.20.Mn, 75.25.Dk, 74.20.Rp, 74.70.Xa}

\maketitle

\section{introduction}
High-temperature cuprate superconductors are realized by carrier doping into antiferromagnetic Mott insulators \cite{keimer15}. The importance of spin fluctuations was widely recognized for the high-$T_{c}$ mechanism \cite{anderson87,scalapino12}, and spin fluctuation spectra were studied intensively in space of energy $\omega$ and momentum $\vq$ by inelastic neutron scattering already slightly after the discovery of high-$T_{c}$ cuprates \cite{thurston89,rossat-mignod91}. Given that it is mobile electrons which form Cooper pairs, charge fluctuations should also be equally important to understand  the high-$T_{c}$ mechanism. 
In particular, the spin-charge stripe order \cite{tranquada95} was observed in La-based cuprates around the hole doping 1/8, where the superconductivity tends to be suppressed \cite{tranquada21}. After the advent of resonance inelastic x-ray scattering (RIXS), charge excitations were investigated in much wider $\vq$-$\omega$ space. There were two major discoveries. One was the charge ordering tendency, which is not accompanied by magnetic order in contrast to the spin-charge stripes, in the other hole-doped cuprates \cite{ghiringhelli12,hashimoto14,peng16,chaix17,arpaia19,yu20,wslee21,lu22,arpaia23} as well as in electron-doped cuprates \cite{da-silva-neto15,da-silva-neto16,da-silva-neto18}. The charge order in electron-doped cuprates was explained in terms of $d$-wave bond-charge order \cite{yamase15b,li17,yamase19}, which is reduced to the electronic nematic order in the limit of zero momentum \cite{yamase00a,yamase00b,metzner00}. The origin of the charge order in the hole-doped cuprates is still controversial \cite{bejas12,allais14,meier14,wang14,atkinson15,yamakawa15,mishra15,zeyher18}. These spin and charge fluctuations were usually studied by focusing on the CuO$_{2}$ plane, namely, in a two-dimensional model with a short-range interaction. 

The other distinguished discovery by RIXS  was charge excitations close to the in-plane momentum $\vq_{\parallel}=(0,0)$. Their origin was discussed controversially in the early days \cite{ishii14,wslee14,greco16,ishii17,dellea17}, but they are now interpreted consistently as low-energy plasmon excitations with a gap at $\vq_{\parallel}=(0,0)$ that is scaled by weak interlayer hopping. In this sense, it may be refereed to as acousticlike plasmons \cite{greco16,hepting18,greco19,greco20,lin20,nag20,fidrysiak21,hepting22,singh22a,hepting23,nag24}.   Therefore, in contrast to usual analyses in a two-dimensional model with a short-range interaction, not only the long-range Coulomb interaction but also the three-dimensional layered structure has to be considered to understand the charge dynamics in cuprates. In particular, it was pointed out theoretically \cite{greco16,greco19} that their $q_{z}$ dispersion is crucial to identify them as acousticlike plasmons. The acousticlike plasmons are realized for a finite $q_{z}$ and become the conventional optical plasmon at $q_{z}=0$ \cite{nuecker89,romberg90,bozovic90}. So far, those RIXS experiments have been performed mainly in single-layer cuprates, one CuO$_{2}$ plane per unit cell. 

A natural question is whether such plasmon excitations can be universal in cuprate superconductors, especially for multilayer systems (multiple CuO$_{2}$ planes in the unit cell). This is particularly interesting because multilayer cuprates can exhibit $T_{c}$ around 100 K, much higher than most of single-layer cuprates. Hence their charge dynamics may contain an important clue to understand the high-$T_{c}$ mechanism.  Recently, the RIXS experiments were performed for the bilayer cuprates ${\rm YBa_{2}Cu_{3}O_{6+\delta}}$ and Ca-doped ${\rm YBa_{2}Cu_{3}O_{6+\delta}}$ \cite{bejas24}. The data suggested a single plasmon mode, similar to that observed in the single-layer cuprates. 

Meanwhile, there were theoretical studies of plasmons in a bilayer system by using the electron-gas model that incorporated the long-range Coulomb interaction (LRC) \cite{fetter74,griffin89}. In this model, electrons are mobile only inside the layer---no hopping between the layers---and electrons interact with each other via the LRC not only inside the layer but also in different layers. Griffin and Pindor \cite{griffin89} performed detailed and clear analysis in the context of cuprates. They predicted two plasmon modes, in-phase and out-of-phase modes between the two layers inside the unit cell, which we may call the $\omega_{+}$ mode and $\omega_{-}$ mode, respectively, following Ref.~\cite{griffin89}. The $\omega_{+}$ mode bears a feature similar to plasmons in the single-layer system and the $\omega_{-}$ mode is a new feature in the bilayer system. Both $\omega_{\pm}$ modes were  predicted to be gapless and the $\omega_{-}$ mode had a negligible $q_{z}$ dependence and energy lower than the $\omega_{+}$ mode \cite{griffin89}. In cuprates, however, it is expected that there is at least sizable intrabilayer hopping. Moreover, it is questionable whether the electron-gas model \cite{fetter74,griffin89} is applicable to materials close to half-filling such as cuprates and nickelates, which are usually regarded as electron-liquid systems---rather the electron-gas model has been applied typically to semiconductor superlattices \cite{jain88}. Nonetheless, the formulae of the LRC obtained in the electron-gas model \cite{fetter74} were frequently employed even for systems far away from the low-electron density limit \cite{griffin89,levallois16,hepting18,boyd22,bejas24}. We infer that its reason may lie in capturing correctly the plasmon continuum specific to layered systems \cite{fetter74,kresin88,griffin89,bill03} and the $q^{-2}$ singularity of the LRC in the limit of $\vq \rightarrow {\bf 0}$, leading to plasmons, and in an expectation that the essential physics for the long-wavelength limit ($\vq \rightarrow {\bf 0}$) may be described in the electron-gas model.  

The theoretical analysis in Ref.~\cite{bejas24} fixed partially an apparent drawback of the electron-gas model by including the lattice structure in the kinetic term and also a finite intrabilayer hopping \cite{grecu73}, yet keeping the same functional form of the LRC as that in the electron-gas model \cite{fetter74,griffin89}. Hence the serious drawbacks remain: i) the interaction term lost $2\pi$ periodicity inside the layer although the lattice structure was considered for the kinetic term, ii) the charge excitation spectrum along the $q_{z}$ direction was tacitly assumed to have $2\pi$ periodicity, which is not correct because the spectral weight loses such periodicity, and iii) RIXS experiments were usually performed in wide $\vq$-$\omega$ space \cite{ishii14,wslee14,ishii17,dellea17,hepting18,lin20,nag20,hepting22,singh22a,hepting23,nag24}, not restricted to the vicinity of $\vq_{\parallel} = (0,0)$ and $q_{z}c=0$; in fact, $q_{z}c$ was frequently taken away from zero in many RIXS experiments \cite{hepting18,lin20,nag20,hepting22,singh22a,hepting23,nag24}. These features indicates some limitation of previous theoretical analyses \cite{griffin89,bejas24}, which eventually may not be sustainable to a sound comparison with RIXS data especially for materials near half-filling. This is particularly serious since the charge dynamics in multi-layer cuprates, which exhibit $T_{c}$ around 100 K, contains potentially some important hints to the understanding of the high-temperature superconducting mechanism. 

Motivated by this possible importance, we shall derive formulae of the LRC beyond the ones obtained in the electron-gas model by Fetter 50 years ago \cite{fetter74}---we refer to it as the Fetter model---so that the we can apply the formulae to a lattice system with any electron density.  This was already achieved for the single-layer lattice model \cite{becca96,greco16,fidrysiak21}, but is largely unknown for multilayer electron systems. We shall consider a bilayer lattice model as the simplest case. But even in this simplest case, we are not aware of literature to develop the Fetter model. Such a development will eventually make it possible to achieve a detailed comparison between theory and experiments even far away from the low-electron density limit, especially for materials modeled by electron liquids such as cuprates, nickelates and others. Moreover, it may be historically important to construct a theory with the LRC applicable to any materials, independent of the electron density, by respecting the full lattice symmetry, for example in graphite \cite{shyu00,anderson19}. 

In this paper, we derive formulae of the LRC on a bilayer lattice by solving the Poisson equation and demonstrate its validity by checking various limits. Our obtained formulae are general and can be applicable to any bilayer  square lattice systems, especially when one wishes to study the charge dynamics in the presence of the LRC; the extension to other lattice geometry is straightforward. Given the current interests in the charge dynamics in cuprates, we choose model parameters appropriate to cuprates, although strong correlation effects specific to cuprates are not taken into account. Rather we give weight to  clarifying the characteristic feature of the charge dynamics in the bilayer lattice model with the LRC. The inclusion of the strong electron correlations in the context of the $t$-$J$-$V$ model \cite{greco16} and the further extension to three- and four-layer systems are left in the future. 

The present manuscript is organized as follows. In Sec.~II, we introduce a bilayer model with a density-density interaction. We then study the case of the LRC in Sec.~II~B. Equations~(\ref{Vq}) and (\ref{Vqp}) are the formulae beyond the Fetter model---one of the most important results in the present work. The analytical form of the dynamical charge susceptibility is presented in the non-interacting case, the bilayer case only with the intrabilayer hopping, and the most general case. Numerical results are presented in Sec.~III. First we turn off the interbilayer hopping $t_{z}^{'}$, clarify the overall charge-excitation spectrum, and elaborate the property of the $\omega_{\pm}$ modes. Next we introduce $t_{z}^{'}$ and show its major effect. In the discussion section (Sec.~IV), we make a comparison with RIXS data, the Fetter model, and the case of a short-range interaction, and discuss a relation to charge ordering tendencies in cuprates. Conclusions are given in Sec.~V. We also study the limit of the single-layer case in our formulae [Eqs.~(\ref{Vq}) and (\ref{Vqp})] in Appendix A and clarify a fine structure of the $\omega_{\pm}$ modes at $q_{z}c=\pi$ in Appendix B.

\section{Formalism} 
\subsection{Hamiltonian}
\begin{figure}[th]
\centering
\includegraphics[width=8cm]{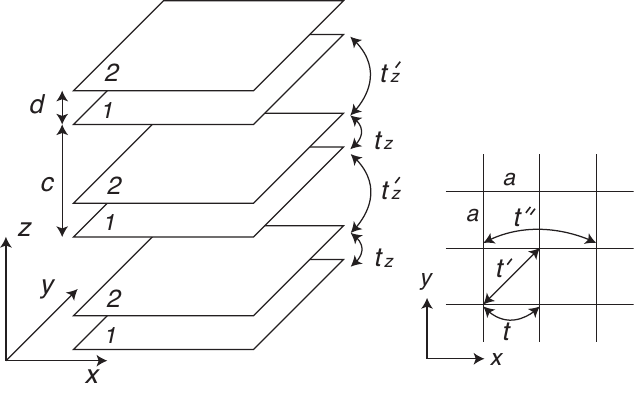}
\caption{Bilayer model. Each layer forms a square lattice and the hopping integrals are considered up to the third nearest-neighbor sites, $t$, $t'$, and $t''$ (right figure). The unit cell contains two layers, 1 and 2. The intrabilayer hopping is given by $t_{z}$ and the interbilayer hopping is $t_{z}'$. The lattice constants are given by $a$, $a$ and $c$ along the $x$, $y$, and $z$ directions, respectively, and the intrabilayer distance is $d (\leq c/2)$. 
}
\label{bilayer}
\end{figure}

We consider an electron system interacting with the LRC on a bilayer square lattice stacked along the $z$ axis with the intrabilayer distance $d$ and interbilayer distance $c$ as shown in \fig{bilayer}; we can assume $0< d \leq c/2$ without losing generality. 
%%%%%%%%%%%%%%%%%%%%% FIG. 1 %%%%%%%%%%%%%%%%%%%%%%%%
%%%%%%%%%%%%%%%%%%%%%%%%%%%%%%%%%%%%%%%%%%%%%%%%%
The resulting Hamiltonian consists of two terms 
\be
\mathcal{H}=\mathcal{H}_{0}+\mathcal{H}_{I} \,.
\ee
$\mathcal{H}_{0}$ describes the kinetic term and is given by 
\be
\mathcal{H}_{0}= \sum_{\vk, \sigma}  
\begin{pmatrix} 
c_{1 \vk \sigma}^{\dagger}   c_{2 \vk \sigma}^{\dagger} 
\end{pmatrix}
\begin{pmatrix}
\xi_{\vk} & \varepsilon_{\vk} {\rm e}^{i k_{z} d} \\
\varepsilon_{\vk}^{*} {\rm e}^{-i k_{z} d} & \xi_{\vk}
\end{pmatrix}
\begin{pmatrix}
c_{1 \vk \sigma} \\
c_{2 \vk \sigma}
\end{pmatrix} \,,
\label{H0}
\ee
where $c_{1 \vk \sigma}^{\dagger}$ ($c_{2 \vk \sigma}^{\dagger}$) and $c_{1 \vk \sigma}$ ($c_{2 \vk \sigma}$) are creation and annihilation operators of electrons with momentum $\vk$ and spin $\sigma$ on the layer 1 (2);  $\xi_{\vk}$ and $\varepsilon_{\vk}$ are the in-plane and out-of-plane dispersions, respectively, 
and are given by 
\bea
&&\xi_{\vk} = -2 t (\cos k_{x}a + \cos k_{y}a ) -4 t' \cos k_{x}a \cos k_{y}a  
-2t'' (\cos 2k_{x}a + \cos 2k_{y}a ) - \mu \, , 
\label{xiplane} \\
&& \varepsilon_{\vk} = \varepsilon_{\vk}^{\perp} +  \varepsilon_{\vk}^{\perp '} {\rm e}^{-i k_{z} c} \, .
\label{xiperp}
\eea
The hopping integrals $t_{z}$ and $t_{z}'$ between the layers (see \fig{bilayer}) yield the dispersions $ \varepsilon_{\vk}^{\perp}$ and $ \varepsilon_{\vk}^{\perp '}$, respectively, 
\bea
&&\varepsilon_{\vk}^{\perp}= - t_{z} (\cos k_{x}a - \cos k_{y}a)^{2} \, , \\
&&\varepsilon_{\vk}^{\perp '}= - t_{z}^{'} (\cos k_{x}a - \cos k_{y}a)^{2} \, .
\eea
The form factor $\cos k_{x}a - \cos k_{y}a$ is expected in local density approximation (LDA) band calculations \cite{andersen95}. 

The lattice sites in the layer 1 in \fig{bilayer} are specified by $\vr_{i}= n_{x} a \hat{{\bf x}} + n_{y} a \hat{{\bf y}}  +  n_{z} c \hat{{\bf z}}$ with  the unit vectors $\hat{{\bf x}}$, $\hat{{\bf y}}$, and $\hat{{\bf z}}$ along the $x$, $y$, and $z$ directions, respectively, and with integers $n_{x}$, $n_{y}$, and $n_{z}$. The lattice sites on the layer 2 are then described by $\vr_{i}+{\bf d}$ with ${\bf d}=(0,0,d)$. Hence in real space, the density-density interaction is given by 
\bea
&&H_{I}=\frac{1}{2} \sum_{i j} \left[ n_{1}(\vr_{i}) V( \vr_{j} - \vr_{i}) n_{1}(\vr_{j})  \right. \nonumber \\ 
&&\hspace{20mm} + \left. n_{2}(\vr_{i}+{\bf d}) V( \vr_{j} - \vr_{i}) n_{2}(\vr_{j} +{\bf d})  \right.  \nonumber \\
&& \hspace{20mm} + \left. n_{1}(\vr_{i}) V( \vr_{j} + {\bf d} - \vr_{i}) n_{2}(\vr_{j} +{\bf d}) \right.   \nonumber \\
&& \hspace{20mm}  + \left. n_{2}(\vr_{i}+{\bf d}) V( \vr_{j} - \vr_{i} - {\bf d}) n_{1}(\vr_{j}) \right] \,. 
\eea
After the Fourier transform, 
\bea
&& n_{1}(\vr_{i}) =  \frac{1}{N} \sum_{\vq} n_{1}(\vq) {\rm e}^{i \vq \cdot \vr_{i}} \,, \\
&& n_{2}(\vr_{i}+{\bf d}) = \frac{1}{N} \sum_{\vq} n_{2}(\vq) {\rm e}^{i \vq \cdot (\vr_{i} +{\bf d}) } \,, 
\eea
we obtain 
\be
\mathcal{H}_{I} = \frac{1}{2N} \sum_{\vq} 
\begin{pmatrix}
n_{1}(\vq) \; n_{2}(\vq) 
\end{pmatrix}
\begin{pmatrix}
V(\vq) \; V^{'}(\vq) \\
V^{'}(-\vq)  \; V(\vq) 
\end{pmatrix}
\begin{pmatrix}
n_{1}(-\vq)  \\
n_{2}(-\vq) 
\end{pmatrix} \,,
\label{HI}
\ee
where $N$ is the total number of lattice sites; $n_{1}(\vq)= \sum_{\vk, \sigma} c_{1 \vk  \sigma}^{\dagger}  c_{1 \vk + \vq \sigma}$ and $n_{2}(\vq)= \sum_{\vk, \sigma} c_{2 \vk \sigma}^{\dagger}  c_{2 \vk+\vq \sigma}$ are the electron density operators of the momentum $\vq$ in each layer; the Fourier transform of the interaction part is given by 
\bea
 &&V({\bf l}) = \frac{1}{N} \sum_{\vq} V(\vq) {\rm e}^{i \vq \cdot {\bf l}} 
 \label{vq} \,,\\
 &&V({\bf l}+ {\bf d}) = \frac{1}{N} \sum_{\vq} V^{'}(\vq) {\rm e}^{i \vq \cdot ({\bf l} + {\bf d})} 
 \label{v'q} \, ,
 \eea
with ${\bf l} = \vr_{j} -\vr_{i}$.

\subsection{LRC on bilayer lattice}
Equation~(\ref{HI}) is applicable to any density-density interaction. In particular,  we consider the LRC in the present study. See Sec.~IV~C for the case of a short-range interaction, which is much simpler than the case of the LRC. 

Needless to say, the functional factor of $V({\bf l})$ corresponds to the well-known LRC $V({\bf l}) \propto  \frac{e^{2}}{| {\bf l} |}$ in continuum space and $e$ is the electric charge; its Fourier transform is given by $\sim \frac{e^{2}}{q^{2}}$. If one is interested in the vicinity of momentum $\vq={\bf 0}$, such a functional form could be justified to be employed even on a lattice system; at least, the singularity of the LRC is correctly captured. However, the asymptotic form in a bilayer system is not correctly reproduced as we point it out in Sec.~IV~B. Moreover, typically RIXS measurements are performed at $\vq$ away from zero to avoid a very strong elastic signal that masks the low-energy charge excitations \cite{ishii14,wslee14,ishii17,dellea17,hepting18,lin20,nag20,hepting22,singh22a,hepting23,nag24}. Therefore, it is requisite to use the LRC that respects the lattice structure. It is already known how to compute the LRC on a Bravais lattice \cite{becca96}, but we are not aware of literature to provide expressions of $V(\vq)$ and $V'(\vq)$ for the bilayer, more generally multilayer, case beyond the 50-year-old model by Fetter \cite{fetter74,griffin89}. In fact, we have recognized that the derivation of the LRC on a multilayer system is not straightforward at all. We therefore focus on the bilayer case in the present work and elucidate characteristic features of the charge dynamics in the presence of the LRC, which contains enough insights beyond the available knowledge \cite{griffin89}. 

To obtain the LRC on a lattice, the standard procedure is to solve the Poisson  equation as demonstrated in Ref.~\cite{becca96} for the single-layer case. We start from the Poisson equation with spatial anisotropy: 
\be
\epsilon_{\parallel} \left(  
\frac{\partial^{2}}{\partial x^{2}} +  \frac{\partial^{2}}{\partial y^{2}} \right) G(\vr) +
\epsilon_{\perp} \frac{\partial^{2}}{\partial z^{2}}G(\vr)  = -\delta(\vr) \,,
\label{Poisson}
\ee
where $\epsilon_{\parallel} (\epsilon_{\perp})$ is the dielectric constant parallel (perpendicular) to the layer. On the lattice, the second differential is replaced by the difference 
\bea
&&\frac{\epsilon_{\parallel}}{a^{2}} \left\{ \left[ 
G(\vr_{i} + a \hat{\bf x}) - 2 G(\vr_{i}) +  G(\vr_{i} - a \hat{\bf x})  \right] \right. \nonumber \\
&&\hspace{5mm} +\left. \left[ G(\vr_{i} + a \hat{\bf y}) - 2 G(\vr_{i}) +  G(\vr_{i} - a \hat{\bf y}) 
 \right] \right\}  \nonumber \\
&&\hspace{0mm} +  \epsilon_{\perp} \left[ h_{1} G(\vr_{i} + c \hat{\bf z}) + h_{2} G(\vr_{i} + d \hat{\bf z}) +
 h_{3} G(\vr_{i}) +  h_{4} G(\vr_{i} - (c-d) \hat{\bf z} ) \right] \nonumber \\
&& \hspace{0mm}  =- \frac{1}{N}\sum_{\vq}{\rm e}^{i \vq \cdot \vr_{i}} \,.
\eea
Here we use 4 points to evaluate $ \frac{\partial^{2} G(\vr)}{ \partial z^{2}}$ so that the equation is valid even for  $d=0$; the coefficients are determined uniquely by the Taylor expansion, yielding 
\bea
&&h_{1}= \frac{2 (c-2d)}{c(2c-d) (c-d)} \,,  \label{h1}\\
&&h_{2}= \frac{2}{c(c-d)} \,, \\ 
&&h_{3}= -\frac{4}{c(c-d)} \,, \\ 
&&h_{4}= \frac{2 (c+d)}{c(2c-d) (c-d)}  \label{h4} \,.
\eea
From the Fourier transform defined in Eqs.~(\ref{vq}) and (\ref{v'q}), we obtain 
\bea
&& \left[ \frac{2 \epsilon_{\parallel}} {a^{2}} (2-\cos q_{x}a - \cos q_{y}a) - \epsilon_{\perp} (h_{1} {\rm e}^{i q_{z} c} + h_{3}) \right] G(\vq) \nonumber \\
&& - \epsilon_{\perp} {\rm e}^{i q_{z}d} ( h_{2} + h_{4} {\rm e}^{-i q_{z}c} ) G^{'}(\vq) = 1 
\label{Poisson1} \,.
\eea
Note that $G^{'}(\vq)$ appears in the second line, suggesting a necessity of two independent equations. The other Poisson equation on a bilayer lattice may be considered at ${\bf r}+ {\bf d}$ in \eq{Poisson}: 
\bea
&&\frac{\epsilon_{\parallel}}{a^{2}} \left\{ \left[ 
G(\vr_{i} + {\bf d} + a \hat{\bf x}) - 2 G(\vr_{i} + {\bf d}) +  G(\vr_{i} + {\bf d} - a \hat{\bf x})
 \right] \right. \nonumber \\
 &&\hspace{5mm} + \left.\left[ 
 G(\vr_{i} + {\bf d} + a \hat{\bf y}) - 2 G(\vr_{i} + {\bf d}) +  G(\vr_{i} + {\bf d} - a \hat{\bf y})
 \right] \right\} \nonumber \\
&&\hspace{0mm} + \epsilon_{\perp} \left[ h_{5} G(\vr_{i} + c \hat{\bf z}) + h_{6} G(\vr_{i} + d \hat{\bf z}) +
 h_{7} G(\vr_{i}) +  h_{8} G(\vr_{i} - (c-d) \hat{\bf z} ) \right]  \nonumber \\
 && =0 \,.
 \label{Possion2}
\eea
The right hand side becomes zero because $\delta(\vr_{i} + {\bf d})=0$ in the condition $0 < d \leq c/2$. When we determine the coefficients $h_{5}, h_{6}, h_{7}, h_{8}$ by the Taylor expansion, we find that $h_{5}=h_{4}$, $h_{6}=h_{3}$, $h_{7}=h_{2}$, and $h_{8}=h_{1}$. We then obtain 
\bea
&& \left[ \frac{2 \epsilon_{\parallel}} {a^{2}} (2-\cos q_{x}a - \cos q_{y}a) 
- \epsilon_{\perp} (h_{1} {\rm e}^{-i q_{z} c} + h_{3}) \right] G^{'}(\vq) \nonumber \\
&& - \epsilon_{\perp} {\rm e}^{-i q_{z}d} ( h_{2} + h_{4} {\rm e}^{i q_{z}c} ) G(\vq) = 0 
\label{Poisson2} \,.
\eea
Equations~(\ref{Poisson1}) and (\ref{Poisson2}) are reduced to a $2 \times 2$ matrix equation
\be
\begin{pmatrix}
A_{11} \; A_{12} \\
A_{12}^{*}  \;  A_{11}^{*} 
\end{pmatrix}
\begin{pmatrix}
G(\vq) \\
G^{'}(\vq) 
\end{pmatrix} 
= 
\begin{pmatrix}
1  \\
0
\end{pmatrix} \,,
\ee
where 
\bea
&&A_{11}= \frac{2 \epsilon_{\parallel}}{a^{2}} (2 - \cos q_{x}a - \cos q_{y}a ) -   
\epsilon_{\perp}(h_{1} {\rm e}^{iq_{z} c} + h_{3}) \,, 
\label{A11} \\
&& A_{12}= -\epsilon_{\perp} {\rm e}^{iq_{z} d} ( h_{2} + h_{4} {\rm e}^{-iq_{z} c} ) \,. 
\label{A12}
\eea
Hence we obtain the solution of the Poisson equation for the bilayer system in momentum space  as 
\bea
&&G(\vq)= \frac{A_{11}^{*}}{ |A_{11}|^2 -  |A_{12}|^2} \,, \\ 
&&G^{'}(\vq)= \frac{-A_{12}^{*}}{ |A_{11}|^2 -  |A_{12}|^2} \,.
\eea

The above solution is related to the LRC via
\be
V(\vq)=\frac{e^{2}}{a^{2}c} {\rm Re} G(\vq)\,, \quad V^{'}(\vq)=\frac{e^{2}}{a^{2}c} G^{'}(\vq)  \,,
\label{Vq-define}
\ee
where the factor $1/(a^{2}c)$ comes form the volume of the unit cell. The diagonal component of $V(\vq)$ is given by the real part of $G(\vq)$. This is because only the even term with respect to $\vq$ contributes to the diagonal part of the interaction in \eq{HI}---the term $\sin q_{z}c$ originating from \eq{A11} is canceled out. 

It may be convenient to present the explicit expressions of $V(\vq)$ and $V^{'}(\vq)$, which are computed as: 
\bea
&& V(\vq)=\frac{V_{c}}{{\rm det}\tilde{V}}\left[
\alpha(2-\cos q_{x}a - \cos q_{y}a) - \frac{1}{2} \tilde{h}_{3} - \frac{1}{2} \tilde{h}_{1} \cos q_{z}c \right]
\label{Vq} \,, \\
&&  V^{'}(\vq)=\frac{1}{2} \frac{V_{c}}{{\rm det}\tilde{V}}\left[ 
\tilde{h}_{2} \cos q_{z}d + \tilde{h}_{4} \cos q_{z}(c-d)  -i \tilde{h}_{2}\sin q_{z}d + i   \tilde{h}_{4}\sin q_{z}(c-d)  \right] 
\label{Vqp}  \,, \\
&& {\rm det}\tilde{V} = \left[ \alpha ( 2 - \cos q_{x}a - \cos q_{y}a) \right]^{2} 
- \alpha ( 2 - \cos q_{x}a - \cos q_{y}a) (\tilde{h}_{1} \cos q_{z}c + \tilde{h}_{3})  \nonumber \\
&& \hspace{20mm} + \frac{6 c^{2}}{(c-d)(2c-d)} (1 - \cos q_{z}c) \,.
\eea
Here $\tilde{h}_{i} = c^{2} h_{i}$ with $i=1,2,3,4$ is a dimensionless factor given in Eqs.~(\ref{h1})--(\ref{h4}); $V_{c}= \frac{e^{2} c}{2 a^{2} \epsilon_{\perp}}$ and $\alpha= \frac{c^{2} \epsilon_{\parallel}}{a^{2}\epsilon_{\perp}}$, which are the same notations as those in the single-layer case \cite{greco19,nag20,greco20,hepting22,zinni23,hepting23,nag24}; $V_{c}$ has the dimension of energy and so do $V(\vq)$ and $V^{'}(\vq)$; $\alpha$ describes the anisotropy between the in-plane direction and the perpendicular direction.

\subsection{Dynamical charge susceptibility} 
As shown in Eqs.~(\ref{H0}) and (\ref{HI}), the Hamiltonian is described by a $2 \times 2$ matrix. Thus the charge susceptibility $\kappa_{ij}(\vq, \omega)$ is also described by a $2 \times 2$ matrix. We compute $\kappa_{i j}$ in the random phase approximation (RPA), yielding 
\be
\kappa_{i j}(\vq, \omega) =  \kappa_{i j}^{0}(\vq, \omega)+  \sum_{l_{1}, l_{2}}\kappa_{i l_{1}}^{0}(\vq, \omega)  V_{l_{1} l_{2}}(\vq) \kappa_{l_{2} j} (\vq, \omega) \, , 
\label{RPA}
\ee
where $i, j, l_{1}, l_{2} = 1$ and $2$; $\kappa^{0}_{i j}(\vq, \omega)$ describes a simple bubble diagram and is computed as 
\bea
&&\kappa_{11}^{0}(\vq, \omega) = \frac{1}{2N} \sum_{\vk} (g_{++} + g_{+-} + g_{-+} + g_{--}) \,, \\
&&\kappa_{12}^{0}(\vq, \omega) = \frac{1}{2N} \sum_{\vk} \frac{\varepsilon_{\vk} \varepsilon_{\vk+\vq}^{*} {\rm e}^{-i q_{z}d}}{ |\varepsilon_{\vk} | |\varepsilon_{\vk + \vq} |} (g_{++} - g_{+-} - g_{-+} + g_{--}) \,, 
\label{ko12}\\
&&\kappa_{21}^{0}(\vq, \omega) =  \frac{1}{2N} \sum_{\vk} \frac{\varepsilon_{\vk}^{*} \varepsilon_{\vk+\vq} {\rm e}^{i q_{z}d}}{ |\varepsilon_{\vk} | |\varepsilon_{\vk + \vq} |} (g_{++} - g_{+-} - g_{-+} + g_{--})  \,, 
\label{ko21} \\
&&\kappa_{22}^{0}(\vq, \omega) =\kappa_{11}^{0}(\vq, \omega) \,, 
\eea
where $\vk$ is defined in the first Brillouin zone and we omit the arguments on the right hand side in $g_{\mu \nu}$ with $\mu, \nu=+$ and $-$, which are given by 
\be
g_{\mu \nu} (\vk, \vq, \omega)= \frac{f(\lambda_{\mu}(\vk)) - f(\lambda_{\nu}(\vk + \vq))} {\lambda_{\mu} (\vk) +\omega + i \Gamma - \lambda_{\nu}(\vk + \vq)} \, .
\label{gmunu}
\ee
$f(x)$ is the Fermi distribution function, the eigenenergies $\lambda_{\pm} (\vk) = \xi_{\vk} \pm | \varepsilon_{\vk} |$ describe the antibonding and bonding bands, respectively [see Eqs.~(\ref{xiplane}) and (\ref{xiperp})], and $\Gamma (>0)$ is an infinitesimally small value, but we shall take a finite value for numerical convenience, which may mimics to some extent broadening effects not included in the RPA. Each component of $V_{l_{1} l_{2}}$ in \eq{RPA} is given by the component of the matrix in \eq{HI}. The total susceptibility is then given by 
\be
\kappa(\vq, \omega)= \frac{1}{2} \sum_{i j} \kappa_{i j}(\vq, \omega)
\label{kqw}
\ee
While \eq{kqw} is a compact form, it may be more insightful to present an explicit form of $\kappa(\vq, \omega)$. 

\subsubsection{Non-interacting case for $t_{z}^{'}=0$} 
We first consider the non-interacting charge susceptibility $\kappa^{0}(\vq, \omega)$: 
\bea
&&\kappa^{0}(\vq, \omega) =\frac{1}{2} \left[
 \kappa^{0}_{11}(\vq,  \omega) + \kappa^{0}_{12}(\vq,  \omega)+ \kappa^{0}_{21}(\vq,  \omega)+\kappa^{0}_{22}(\vq,  \omega) 
 \right] \\
&&
\hspace{15mm} =\cos^{2} \frac{q_{z}d}{2}  \kappa^{0}_{\rm even}(\vq, \omega)+  \sin^{2} \frac{q_{z}d}{2} \kappa^{0}_{\rm odd}(\vq, \omega) \,,
\label{kqw0}
\eea
where
\bea
&& \kappa^{0}_{\rm even}(\vq, \omega)  = \frac{1}{N} \sum_{\vk} (g_{++} + g_{--}) \,,\\
&& \kappa^{0}_{\rm odd}(\vq, \omega)  = \frac{1}{N} \sum_{\vk} (g_{+-} + g_{-+} ) \, ,
\eea
and $\kappa^{0}_{\rm even (odd)}(\vq, \omega)$ is the susceptibility from the intraband (interband) scattering  processes and thus is frequently called the even (odd) mode in a bilayer system. The even and odd modes are selected by choosing $q_{z}d=0$ and $\pi$, respectively. 

\subsubsection{Bilayer model with LRC for $t_{z}^{'}$=0} 
Next we introduce the LRC, which yields both intra- and interbilayer interactions, although the hopping along the $z$ direction is restricted only within the intrabilayer. We then obtain 
\bea
&&\kappa(\vq, \omega) = \frac{1}{\mathfrak{det}} \left[ \cos^{2} \frac{q_{z}d}{2} \kappa^{0}_{\rm even}(\vq, \omega) + \sin^{2} \frac{q_{z}d}{2} \kappa^{0}_{\rm odd}(\vq, \omega)  \right. \nonumber \\
&& \left. \hspace{18mm} - \kappa^{0}_{\rm even}(\vq, \omega)\kappa^{0}_{\rm odd}(\vq, \omega) 
V^{''}(\vq) \right] \,,
\label{kqw1}
\eea
where 
\bea
&&\mathfrak{det} = \left[ 1  - \left( V(\vq) + V_{+}(\vq) \right) \kappa^{0}_{\rm even}(\vq,  \omega) \right] \nonumber  \\
&& \hspace{10mm} \times \left[ 1-  \left(V(\vq) - V_{+}(\vq) \right) \kappa^{0}_{\rm odd}(\vq,  \omega)
\right] \nonumber  \\
&&\hspace{12mm} +\kappa^{0}_{\rm even}(\vq, \omega) \kappa^{0}_{\rm odd}(\vq,  \omega) V_{-}(\vq)^{2} \,. 
\label{kqw1det} \\
&& V^{''}(\vq) = V(\vq) - \frac{V^{'}(\vq) + V^{'}(-\vq)}{2} \,, 
\label{Vpm} \\
&& V_{\pm}(\vq)= \frac{V^{'}(\vq) {\rm e}^{i q_{z}d} \pm V^{'}(-\vq) {\rm e}^{-i q_{z}d}}{2} \,.
\label{Vpp}
\eea
It is insightful to compare \eq{kqw1} with \eq{kqw0}. The charge susceptibility can still be described in terms of  $\kappa^{0}_{\rm even}$ and $\kappa^{0}_{\rm odd}$ in \eq{kqw1}, but in contrast to \eq{kqw0}, they cannot be fully decoupled from each other in general. This is because the LRC yields the interbilayer interaction and thus we have an additional term $\kappa^{0}_{\rm even} \kappa^{0}_{\rm odd} V^{''}(\vq)$ in the numerator and two terms in the denominator that describe couplings of $\kappa^{0}_{\rm even}$ and $\kappa^{0}_{\rm odd}$. Hence the even and odd modes are no longer good quantities to specify the charge excitations. This is in a sharp contrast to the case of a short-range interaction; see Sec.~IV~C. Nonetheless there are exceptions. i) The even mode is well defined at $q_{z}c=0$ [see Eqs.~(\ref{kqw1qzc0-1}) and (\ref{kqw1qzc0-2})] because the $\omega_{-}$ mode vanishes at any $q_{\parallel}$. ii) The odd mode is well defined at specific points $\vq_{\parallel}=(0,0)$ and $q_{z}c=2n\pi$ with $n\ne 0$, where the $\omega_{+}$ mode has zero intensity [see Figs.~\ref{kqw-00}(e), \ref{kqw-tz001}(e), and \ref{kqw-tz025}(e)]. 

The origin of $\mathfrak{det}$ in \eq{kqw1det} is readily understood. The RPA susceptibility \eq{RPA} is written in a $2 \times 2$ matrix form 
\be
\hat{\kappa} = \hat{\kappa}^{0} + \hat{\kappa}^{0} \hat{V} \hat{\kappa}  \,,
\ee
where $\hat{V}$ corresponds to the interaction matrix given in \eq{HI}. We then obtain 
\be
\hat{\kappa} = (1-\hat{\kappa}^{0} \hat{V})^{-1} \hat{\kappa}^{0}  \,. 
\label{kappamatrix}
\ee
$\mathfrak{det}$ in \eq{kqw1det} is given by 
\be
\mathfrak{det} = {\rm det} (1- \hat{\kappa}^{0} \hat{V}) \,.
\label{detk}
\ee

The charge excitation spectrum is studied by Im$\kappa(\vq, \omega)$ in $\vq$-$\omega$ space in \eq{kqw1}. As expected, plasmons can be realized when the condition  $\mathfrak{det}=0$ is fulfilled. Hence we may also analyze \eq{detk} in the present study. For numerical convenience, however, we may study the minimum value of $| \mathfrak{det} |$ because of a possible mixture of the continuum spectrum in actual calculations. 

\subsubsection{Bilayer model with LRC and $t_{z}^{'}$}
As shown in \eq{kqw1}, we can still extract $\kappa^{0}_{\rm even}$ and $\kappa^{0}_{\rm odd}$ without $t_{z}^{'}$, even though they are not good quantities. However, once $t_{z}^{'}$ is included, they are not extracted in a convenient way and we obtain:
\bea
&&  \kappa(\vq, \omega) = \frac{\kappa^{0}_{11}+\kappa^{0}_{\cos} - [(\kappa^{0}_{11})^{2} - (\kappa^{0}_{\cos})^{2} -(\kappa^{0}_{\sin})^{2}] [V(\vq)-V^{'}_{+}(\vq)]}{\mathfrak{det}} \,, 
\label{kqw2} \\
&&\mathfrak{det} = 1 - 2 \kappa^{0}_{11}V(\vq) - 2 \kappa^{0}_{\cos} V_{+}^{'}(\vq) + i 2\kappa^{0}_{\sin} V^{'}_{-}(\vq) \nonumber \\
&& \hspace{10mm} +[ (\kappa^{0}_{11})^{2} - (\kappa^{0}_{\cos})^{2} - (\kappa^{0}_{\sin})^{2} ]
[ V(\vq)^{2} - V^{'}(\vq)V^{'}(-\vq)] \,, 
\label{det2} \\
&& V_{\pm}^{'}(\vq) = \frac{1}{2} [V^{'}(\vq) \pm V^{'}(-\vq)] \,,
\label{Vpmp} 
\eea
where $\mathfrak{det}$ is given by \eq{detk} with a finite $t_{z}^{'}$, and we omit the arguments in $\kappa_{11}^{0}$, $\kappa^{0}_{\cos}$, and $\kappa^{0}_{\sin}$ for brevity. The last expression $V_{\pm}^{'}(\vq)$ should not be mixed up with $V_{\pm}(\vq)$ in \eq{Vpp}. $\kappa^{0}_{\cos}$ and $\kappa^{0}_{\sin}$ originate from the factor $\varepsilon_{\vk}\varepsilon_{\vk+\vq}^{*} {\rm e}^{i q_{z}d}$ and its conjugate in Eqs.~(\ref{ko12}) and (\ref{ko21}), and are given by 
\bea
&&\kappa^{0}_{\cos}(\vq, \omega) = \frac{1}{2N} \sum_{\vk} \cos \varphi_{\vk\, \vk+\vq} (g_{++}-g_{+-}-g_{-+}+g_{--} ) \,, 
\label{cosk} \\
&&\kappa^{0}_{\sin}(\vq, \omega) = \frac{1}{2N} \sum_{\vk} \sin \varphi_{\vk\, \vk+\vq} (g_{++}-g_{+-}-g_{-+}+g_{--} ) \,.
\label{sink}
\eea
Here 
\be
\varphi_{\vk\, \vk+\vq} = \theta_{\vk} - \theta_{\vk+\vq} - q_{z}d
\label{phi2}
\ee
and $\theta_{\vk}$ comes from the hopping along the $z$ direction, 
\bea
&& \varepsilon_{\vk}= | \varepsilon_{\vk} | {\rm e}^{i \theta_{\vk}} \,, \\
&& | \varepsilon_{\vk} | = \sqrt{(\varepsilon_{\vk}^{\perp} + \varepsilon_{\vk}^{\perp'} \cos k_{z}c)^{2} 
+ (\varepsilon_{\vk}^{\perp'} \sin k_{z}c)^{2}} \,.
\eea
Although $\kappa^{0}_{\rm even}$ and $\kappa^{0}_{\rm odd}$ do not appear in \eq{kqw2}, one can check that \eq{kqw2} is reduced to \eq{kqw1} by setting $t_{z}^{'}=0$ as it should be.

\section{Results}
A choice of our model parameters is arbitrary and may depend on a target material. Since we wish to make a comparison with recent experimental data of bilayer cuprate high-temperature superconductors \cite{bejas24}, we choose the following parameter set to achieve a reasonable comparison: $t'/t=-0.30$, $t''/t=0.15$, $d/c=3.36/11.68$ \cite{bejas24}, $n=0.79$, $V_{c}/t=130$, $\alpha=40$,  $\Gamma/t=0.01$, and $T/t=0.01$ in most of cases; $n$ is the electron density. Since we will study $t_{z}$ and $t_{z}^{'}$ dependencies, those values are always specified when we present results. We first study the case of $t_{z}^{'}=0$ (see \fig{bilayer}).  In this case, the electronic motion is restricted within the bilayer, but the electrons interact with each other via the LRC between different layers. The effect of $t_{z}^{'}$ shall be clarified later. We use $t$ as the unit of energy and put $t=1$ below. We also put the in-plane lattice constant $a=1$ for brevity. 

\subsection{Charge excitation spectra with \boldmath{$t_{z}^{'}=0$}} 
\subsubsection{Overall features}
%%%%%%%%%%%%%%%%%%%%% FIG. 2  %%%%%%%%%%%%%%%%%%%%%%%%
\begin{figure}[t]
\centering
\includegraphics[width=8cm]{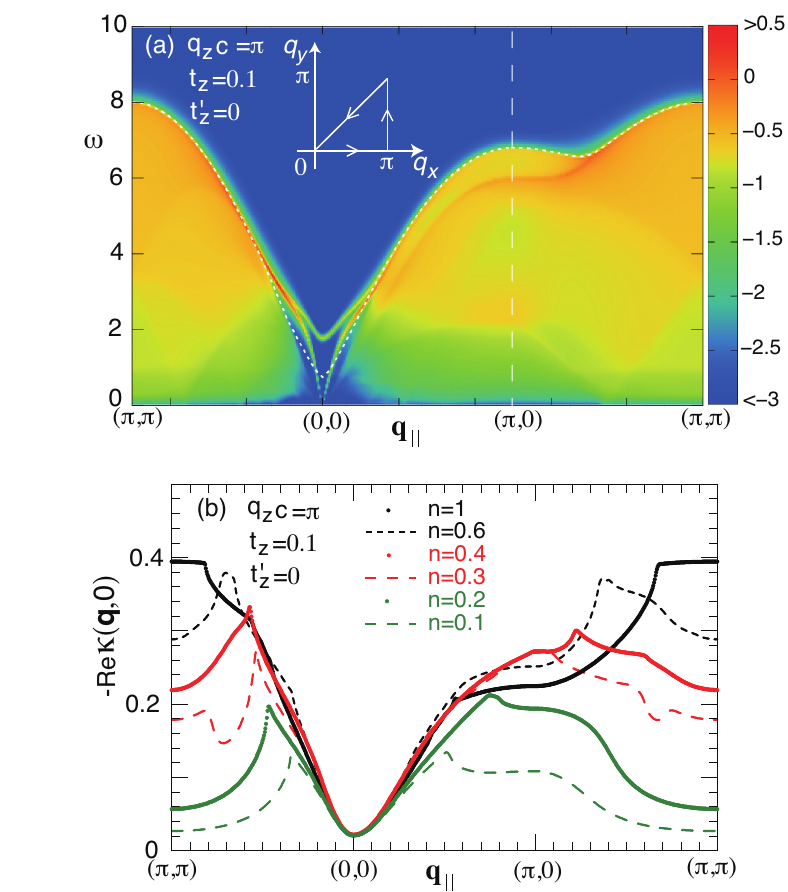}
\caption{(a) Intensity map of $\log_{10} | {\rm Im}\kappa(\vq, \omega)|$ along the symmetry directions shown in the inset; the long-dashed line specifies the position of $(\pi, 0)$. The white dotted curve is the upper boundary of the particle-hole excitations.  Around $\vq_{\parallel}=(0,0)$, there are two collective charge excitation modes, which are plasmons. The lower-energy one becomes gapless at $\vq_{\parallel}=(0,0)$. (b) Real part of the static charge susceptibility along the symmetry directions for several choices of the electron density. 
}
\label{kqw-whole}
\end{figure}
%%%%%%%%%%%%%%%%%%%%%%%%%%%%%%%%%%%%%%%%%%%%%%%%%

Figure~\ref{kqw-whole}(a) is the intensity map of Im$\kappa(\vq, \omega)$ along the symmetry directions $(\pi, \pi)$-$(0,0)$-$(\pi, 0)$-$(\pi, \pi)$ as seen in the inset; $q_{z}c=\pi$ is taken. The white dotted curve is the upper boundary of the particle-hole continuum excitations. Note that low-energy continuum excitations are present even at $\vq_{\parallel}=(0,0)$, because of a finite $t_{z}$. Inside the continuum spectrum, relatively large spectral weight spreads broadly in a high-energy region and the spectral weight tends to be suppressed at lower energy. Above the continuum spectrum around $\vq_{\parallel}=(0,0)$, there are two collective charge modes, which are plasmons.

In \fig{kqw-whole}(b) we plot the real part of the static charge susceptibility for various choices of the electron density, confirming no charge ordering tendency. Although a peak structure forms along the $(0,0)$-$(\pi, \pi)$ direction especially at low density, the susceptibility itself becomes smaller with decreasing the density---it cannot be associated with a potential charge ordering such as a Winger crystallization \cite{ziman} in the present bilayer square lattice system. 

\subsubsection{Maps of the spectral weight}
Having made sure of no charge ordering in the present model, we focus on collective charge excitations around $\vq_{\parallel}=(0,0)$ realized above the upper boundary of the particle-hole excitations in \fig{kqw-whole}. These are plasmon excitations characterized by two modes. Following Ref.~\cite{griffin89}, we may call the upper and lower branches as the $\omega_{+}$ mode and $\omega_{-}$ mode, respectively, although we will show later that the energy of the $\omega_{\pm}$ modes is interchanged for a large $t_{z}$. 
 
 %%%%%%%%%%%%%%%%%%%%% FIG. 3 %%%%%%%%%%%%%%%%%%%%%%%%
\begin{figure}[t]
\centering
\includegraphics[width=16cm]{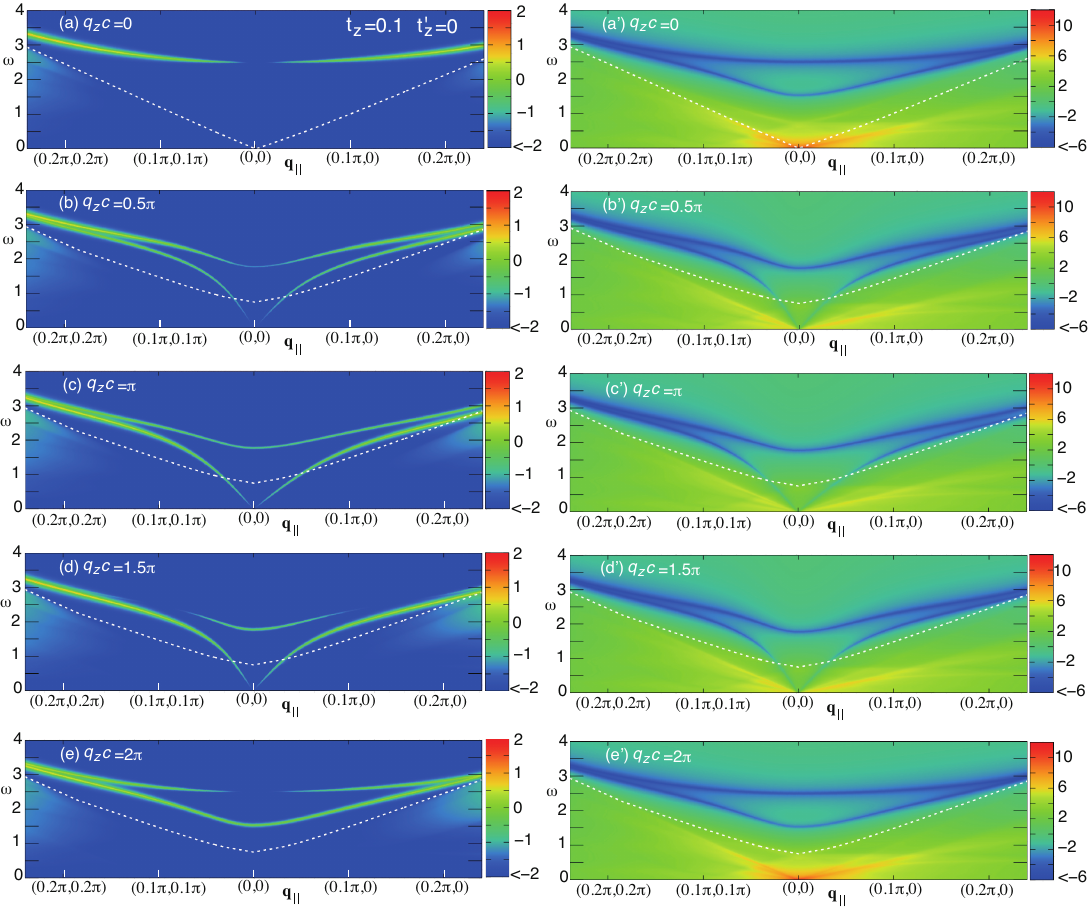}
\caption{Intensity map of $\log_{10} | {\rm Im}\kappa(\vq, \omega)|$ (left panels) and $\log_{10}|\mathfrak{det}|^{2}$ given in \eq{kqw1} (right panels) for a sequence of $q_{z}c$ around a region of $\vq_{\parallel}=(0,0)$: (a) and (a') $q_{z}c=0$, (b) and (b') $q_{z}c=0.5\pi$, (c) and (c') $q_{z}c=\pi$, (d) and (d') $q_{z}c=1.5\pi$, and (e) and (e') $q_{z}c=2\pi$. The white dotted curve denotes the upper boundary of the particle-hole continuum. It goes to zero at $\vq_{\parallel} =(0,0)$ and $q_{z}c=0$ even for a finite $t_{z}=0.1$. To keep an appropriate contrast, the same color is used below -2 and -6 in the left panels and the right panels, respectively. 
}
\label{kqw-00}
\end{figure}
%%%%%%%%%%%%%%%%%%%%%%%%%%%%%%%%%%%%%%%%%%%%%%%%%

Figures~\ref{kqw-00}(a)--(e) show $\vq$-$\omega$ maps focusing on plasmon excitations around $\vq_{\parallel}=(0,0)$ for a sequence of $q_{z}c$. The square of the modulus of the denominator of \eq{kqw1}, namely \eq{detk} is also shown on the right hand side of \fig{kqw-00}. At $q_{z}c=0$ [\fig{kqw-00}(a)], there exists the $\omega_{+}$ mode only---the optical plasmon mode---and no $\omega_{-}$ mode is realized. This is easily understood. Equations~(\ref{Vpm}) and (\ref{Vpp}) show at $q_{z}c=0$ 
\bea
&& V^{''}(\vq) = V(\vq) - V^{'}(\vq) \,, \\
&& V_{+}(\vq) = V^{'}(\vq)\,, \\
&& V_{-}(\vq)=0 \,.
\eea
From \eq{kqw1} we then obtain 
\bea
&& \kappa(\vq, \omega) = \frac{\kappa^{0}_{\rm even}(\vq, \omega) [1- \kappa^{0}_{\rm odd} (\vq, \omega) V^{''}(\vq) ]} {\left[1-\kappa^{0}_{\rm even}(\vq, \omega) (V(\vq)+ V^{'}(\vq)) \right]
 \left[ 1-\kappa^{0}_{\rm odd}(\vq, \omega) (V(\vq) - V^{'}(\vq)) \right] } \,, 
 \label{kqw1qzc0-1}\\
 &&\hspace{10mm} = \frac{\kappa^{0}_{{\rm even}}(\vq, \omega)} 
 {1-\kappa^{0}_{\rm even}(\vq, \omega) (V(\vq)+ V^{'}(\vq))} 
 \label{kqw1qzc0-2} \,.
\eea
That is, the $\omega_{+}$ mode corresponds to the even mode at $q_{z}c=0$. Since $\kappa^{0}_{\rm even} =0$ at $\vq_{\parallel}=(0,0)$ and  $q_{z}c=0$ for finite $\omega$, the intensity of the $\omega_{+}$ mode vanishes at $\vq_{\parallel}=(0,0)$. Physically this  implies that the $\omega_{+}$ mode is in-phase inside the bilayer and thus cannot have spectral weight at $\vq_{\parallel}=(0,0)$ and $q_{z}c=0$ because of the charge conservation---the other mode, namely the $\omega_{-}$ mode, therefore, should be out-of-phase charge-excitation mode in the bilayer system. This assignment is consistent with Ref.~\cite{griffin89}. However, this argument is valid only at $q_{z}c=0$. The relative motion of electrons between the two layers inside the unit cell becomes less trivial as going away from $q_{z}c=0$. In fact, mathematically the in-phase or out-of-phase motion can be confirmed at $q_{z}c=0$ by studying the eigenvector of the susceptibility matrix [\eq{kappamatrix}] at $\vq_{\parallel}$ and $\omega$ where \eq{detk} vanishes and such a result changes as varying $q_{z}c$ from zero. 

Although there is no spectral weight at $\vq_{\parallel}=(0,0)$ and $q_{z}c=0$ at any $t_{z}$, the presence of $t_{z}$ yields low-energy spectral weight at $\vq_{\parallel}=(0,0)$ at $q_{z}c \ne 0$ and the upper boundary of the continuum spectrum has a finite energy.  Above the continuum, we have two modes: the $\omega_{+}$ mode and the $\omega_{-}$ mode for a higher- and lower-energy branch, respectively \cite{griffin89}, as shown in Figs.~\ref{kqw-00}(b)--(d). These modes should not be confused with the even and odd modes specific to a bilayer system, because the LRC yields the interaction among all layers as seen from \eq{kqw1} and the even and odd modes couple to each other. In particular, the $\omega_{-}$ mode shows the gapless dispersion, which may be clear when we recognize that the $\omega_{-}$ mode extends into the continuum spectrum around $\vq_{\parallel}=(0,0)$. At $q_{z}c = 2n\pi$ in \fig{kqw-00}(e), however, the $\omega_{-}$ mode becomes gapped. The $\omega_{+}$ mode loses the spectral weight at $\vq_{\parallel}=(0,0)$, which is due to the property $\kappa^{0}_{\rm even} =0$ there at $q_{z}c=2n\pi$.  A comparison between Figs.~\ref{kqw-00}(a) and (e) highlights that the plasmon excitations at $q_{z}c=0$ are not generic. Rather those at $q_{z}c = 2n\pi$ with $n\ne 0$ are generic: both $\omega_{\pm}$ modes are present and the intensity of the $\omega_{+}$ mode vanishes at $\vq_{\parallel}=(0,0)$ whereas it does not for the $\omega_{-}$ mode.

%%%%%%%%%%%%%%%%%%%%% FIG. 4 %%%%%%%%%%%%%%%%%%%%%%%%
\begin{figure}[t]
\centering
\includegraphics[width=7.5cm]{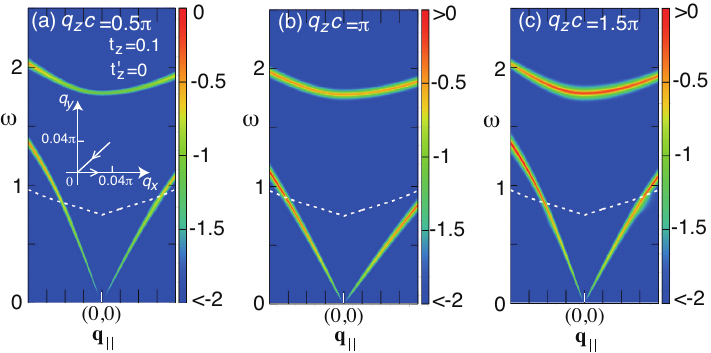}
\caption{Intensity maps of $\log_{10} | {\rm Im}\kappa(\vq, \omega)|$  in the vicinity of $\vq_{\parallel}=(0,0)$ along $(0.04\pi, 0.04\pi)$-$(0, 0)$-$(0.04\pi, 0)$ [see the inset in (a)] for three choices of $q_{z}c$: (a) $q_{z}c=0.5\pi$, (b) $q_{z}c=\pi$, and (c) $q_{z}c=1.5\pi$. The $\omega_{-}$ mode is gapless at $\vq_{\parallel}=(0,0)$ in spite of the presence of $t_{z}=0.1$. The white dotted curve denotes the upper boundary of the particle-hole continuum. To keep an appropriate contrast, the same color is used below -2 and above 0. 
}
\label{kqw-000}
\end{figure}
%%%%%%%%%%%%%%%%%%%%%%%%%%%%%%%%%%%%%%%%%%%%%%%%%

The gapless feature of the $\omega_{-}$ mode is highlighted in Figs.~\ref{kqw-000}(a)--(c) by focusing on the vicinity of $\vq_{\parallel}=(0,0)$. The intensity of the $\omega_{-}$ mode is weaker with decreasing $\omega$ and vanishes at $\omega=0$. An interesting feature is that the $\omega_{-}$ mode crosses the continuum smoothly without showing a clear structure and extends down to zero energy. A reason for that lies in very small spectral weight of the continuum around $\vq_{\parallel}=(0,0)$. In fact, the difference of the spectral weight across the upper boundary of the continuum is not visible in the color scale in \fig{kqw-000}. On the other hand, the $\omega_{+}$ mode has finite spectral weight at $\vq_{\parallel}=(0,0)$ when $0 < q_{z}c < 2 \pi$.

The right panels in \fig{kqw-00} show maps of the denominator of Im$\kappa(\vq, \omega)$ [\eq{kqw1det}], more precisely the square of the modulus of \eq{detk} for  a sequence of $q_{z}c$. The corresponding Im$\kappa(\vq, \omega)$ shown in the left panels can have a peak along (approximately along) the minimum of the denominator of Im$\kappa(\vq, \omega)$ above (below) the continuum. In \fig{kqw-00}(a'), it is interesting that we have two minima of the denominator for a given $\vq_{\parallel}$, but the finite spectral weight is realized only along the $\omega_{+}$ mode and zero along a possible $\omega_{-}$ mode. At $q_{z}c=2\pi$, on the other hand, both $\omega_{+}$ and $\omega_{-}$ modes are realized along the minimum shown in \fig{kqw-00}(e') and have the finite spectral weight except for the $\omega_{+}$ mode at $\vq_{\parallel}=(0,0)$ in \fig{kqw-00}(e).  In \fig{kqw-00}(d), the $\omega_{+}$ mode has strong intensity only near $\vq_{\parallel}=(0,0)$ and the intensity is suppressed as going away from $\vq_{\parallel}=(0,0)$ although the denominator of \eq{kqw1} has a minimum there [\fig{kqw-00}(d')]. This is simply because of the $\vq_{\parallel}$ dependence of the numerator of \eq{kqw1}. The $\omega_{-}$ mode is well visible even inside the continuum---below the white dotted curve in Figs.~\ref{kqw-00}(b)--(d) and Figs.~\ref{kqw-000}(a)--(c)---but the modulus of \eq{detk} gains additional values because of the mixture of the particle-hole excitations [Figs.~\ref{kqw-00}(b')--(d')]. Therefore the peak position of the Im$\kappa(\vq, \omega)$ deviates slightly from the minimum of \eq{detk} below the white dotted curve in Figs.~\ref{kqw-00}(b)--(d) and \ref{kqw-000}(a)--(c).

\subsubsection{A smaller value of $t_{z}$}
We have shown results at $t_{z}=0.1$ and expect qualitatively similar results such as the vanishing of the $\omega_{-}$ mode at $q_{z}c=0$ and the presence of the $\omega_{\pm}$ mode for $q_{z}c\ne 0$ for other choices of $t_{z}$. However, the spectral weight distribution may look different for a smaller $t_{z}$ and it may be worth presenting those results, because several literature \cite{fetter74,griffin89} assumed $t_{z}=0$ in the bilayer system, keeping the LRC among different layers. 

 %%%%%%%%%%%%%%%%%%%%% FIG. 5 %%%%%%%%%%%%%%%%%%%%%%%%
\begin{figure}[th]
\centering
\includegraphics[width=8cm]{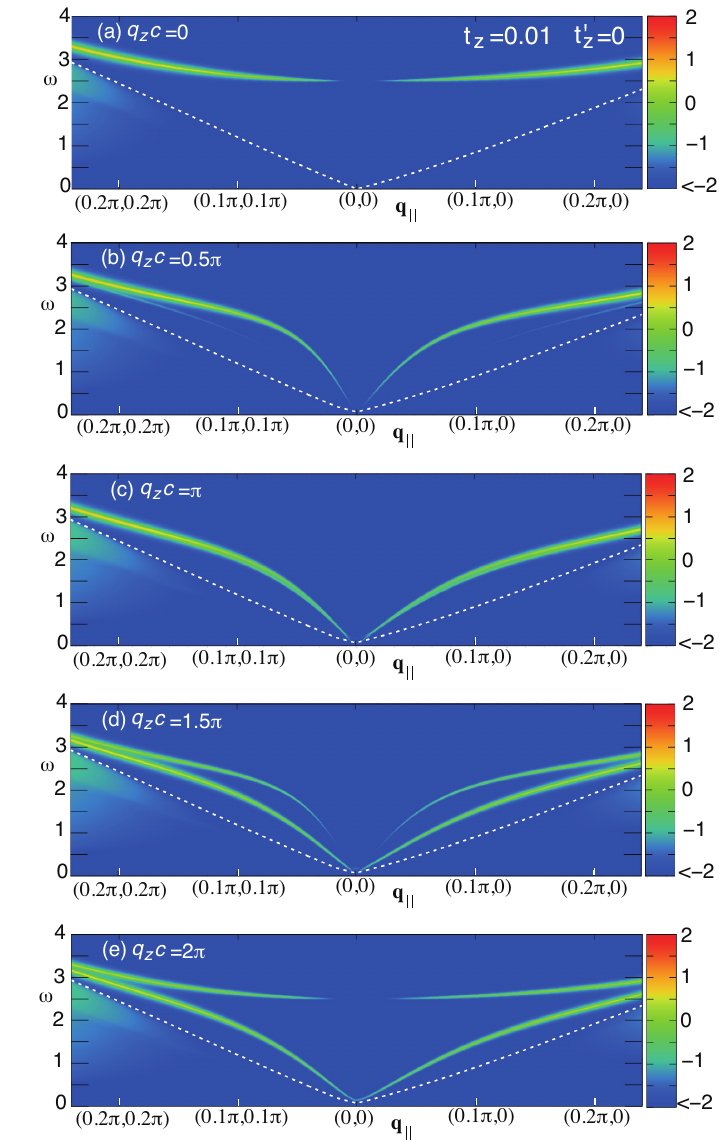}
\caption{Intensity map of $\log_{10} | {\rm Im}\kappa(\vq, \omega)|$ at $t_{z}=0.01$ for a sequence of $q_{z}c$ around a region of $\vq_{\parallel}=(0,0)$: (a) $q_{z}c=0$, (b) $q_{z}c=0.5\pi$, (c) $q_{z}c=\pi$, (d) $q_{z}c=1.5\pi$, and (e) $q_{z}c=2\pi$. The white dotted curve denotes the upper boundary of the particle-hole continuum. It becomes zero at $\vq_{\parallel}=(0,0)$ and $q_{z}c=0$ and finite for the other values of $q_{z}c$. To keep an appropriate contrast, the same color is used below -2. 
}
\label{kqw-tz001}
\end{figure}
%%%%%%%%%%%%%%%%%%%%%%%%%%%%%%%%%%%%%%%%%%%%%%%%%

\begin{figure}[t]
\centering
\includegraphics[width=8cm]{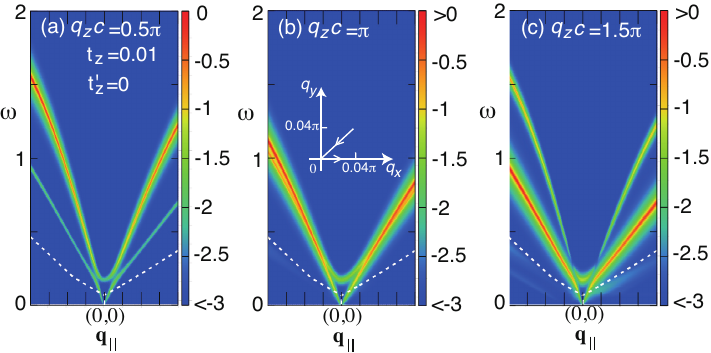}
\caption{Intensity maps of $\log_{10} | {\rm Im}\kappa(\vq, \omega)|$  in the vicinity of $\vq_{\parallel}=(0,0)$ along $(0.04\pi, 0.04\pi)$-$(0,0)$-$(0.04\pi, 0)$ [see the inset in (b)] for (a) $q_{z}c=0.5\pi$, (b) $q_{z}c=\pi$, and (c) $q_{z}c=1.5\pi$. The $\omega_{+}$ mode acquires a small gap at $\vq_{\parallel}=(0,0)$ due to the presence of a small $t_{z}=0.01$. The white dotted curve denotes the upper boundary of the particle-hole continuum. To keep an appropriate contrast, the same color is used below -3 and above 0. 
}
\label{kqw-tz001detail}
\end{figure}

Figures~\ref{kqw-tz001}(a)--(e) are results at $t_{z}=0.01$, which may be compared with Figs.~\ref{kqw-00}(a)--(e). At $q_{z}c=0$ [\fig{kqw-tz001}(a)], the only $\omega_{+}$ mode is realized and quantitative changes are not visible even by employing a small $t_{z}$. However, at $q_{z}c=0.5\pi$ [\fig{kqw-tz001}(b)], it might seem that only one mode is realized. In reality, the $\omega_{+}$ mode has much larger spectral weight and the $\omega_{-}$ mode is hardly visible in the scale of \fig{kqw-tz001}(b). A sharp contrast to \fig{kqw-00}(b) is that the $\omega_{+}$ mode suggests a gapless mode. However, it is not the case. The $\omega_{+}$ mode acquires a small gap due to a small $t_{z}$ whereas the $\omega_{-}$ mode is gapless similar to \fig{kqw-00}(b). These features are checked by magnifying a region around $\vq_{\parallel}=(0,0)$ as shown in \fig{kqw-tz001detail}(a).

%%%%%%%%%%%%%%%%%%%%% FIG. 6 %%%%%%%%%%%%%%%%%%%%%%%%
%%%%%%%%%%%%%%%%%%%%%%%%%%%%%%%%%%%%%%%%%%%%%%%%%

At $q_{z}c=\pi$ [\fig{kqw-tz001}(c)], still only one mode seems to exist, but both $\omega_{\pm}$ modes do exist.  A split is only visible by magnifying the vicinity around $\vq_{\parallel}=(0,0)$ as shown in \fig{kqw-tz001detail}(b) and the $\omega_{+}$ mode seems to almost overlap with the $\omega_{-}$ mode in the other regions. This feature is typical for a small $t_{z}$ at $q_{z}c=\pi$. From \eq{Vpp}, we can check that  $V_{-}(\vq)=0$ at $q_{z}c=\pi$. Equations~(\ref{kqw1}) and (\ref{kqw1det}) then imply that the plasmon dispersion is given by 
\be
\left[ 1- \kappa_{\rm even}^{0}(V(\vq) + V_{+}(\vq)) \right]
\left[ 1- \kappa_{\rm odd}^{0}(V(\vq )- V_{+}(\vq)) \right] =0 \,.
\label{split-eq}
\ee
When $t_{z}$ is small, we have $\kappa_{\rm even}^{0}\approx  \kappa_{\rm odd}^{0}$. Equation~(\ref{Vpp}) indicates that $V_{+}(\vq) \propto  \tilde{h}_{2} + \tilde{h}_{4}\cos q_{z}c =   \tilde{h}_{2} - \tilde{h}_{4} (>0)$ at $q_{z}c=\pi$ for $0 < d < c/2$. This means that $V_{+}(\vq)$ has a minimum at $q_{z}c=\pi$ because $\tilde{h}_{2}$ and $\tilde{h}_{4}$ are positive. This analysis explains the reason why the $\omega_{\pm}$ modes become almost degenerate at  $q_{z}c=\pi$. To see the splitting of these two modes at $q_{z}c=\pi$ more clearly even away from $\vq_{\parallel}=(0,0)$, the damping factor $\Gamma$ should be taken smaller; see Appendix~B.

At $q_{z}c=1.5\pi$ [\fig{kqw-tz001}(d)], the $\omega_{+}$ mode is well separated, although its spectral weight is relatively weak around $\vq_{\parallel}=(0,0)$. When we magnify such a region in \fig{kqw-tz001detail}(c), it is clearly visible that the $\omega_{+}$ mode is acousticlike with a small gap at $\vq_{\parallel}=(0,0)$ due to a small $t_{z}$ whereas the $\omega_{-}$ mode is gapless. 

At $q_{z}c=2\pi$ [\fig{kqw-tz001}(e)], the $\omega_{\pm}$ modes are well separated---the $\omega_{+}$ mode is opticallike and the $\omega_{-}$ mode is acousticlike with a small gap coming from a small $t_{z}$. This is in a sharp contrast to the cases for $q_{z}c \ne 2n\pi$ [Figs.~\ref{kqw-tz001}(b)--(d) and \ref{kqw-tz001detail}(a)--(c)].

%%%%%%%%%%%%%%%%%%%%% FIG. 7 %%%%%%%%%%%%%%%%%%%%%%%%
\begin{figure}[t]
\centering
\includegraphics[width=8cm]{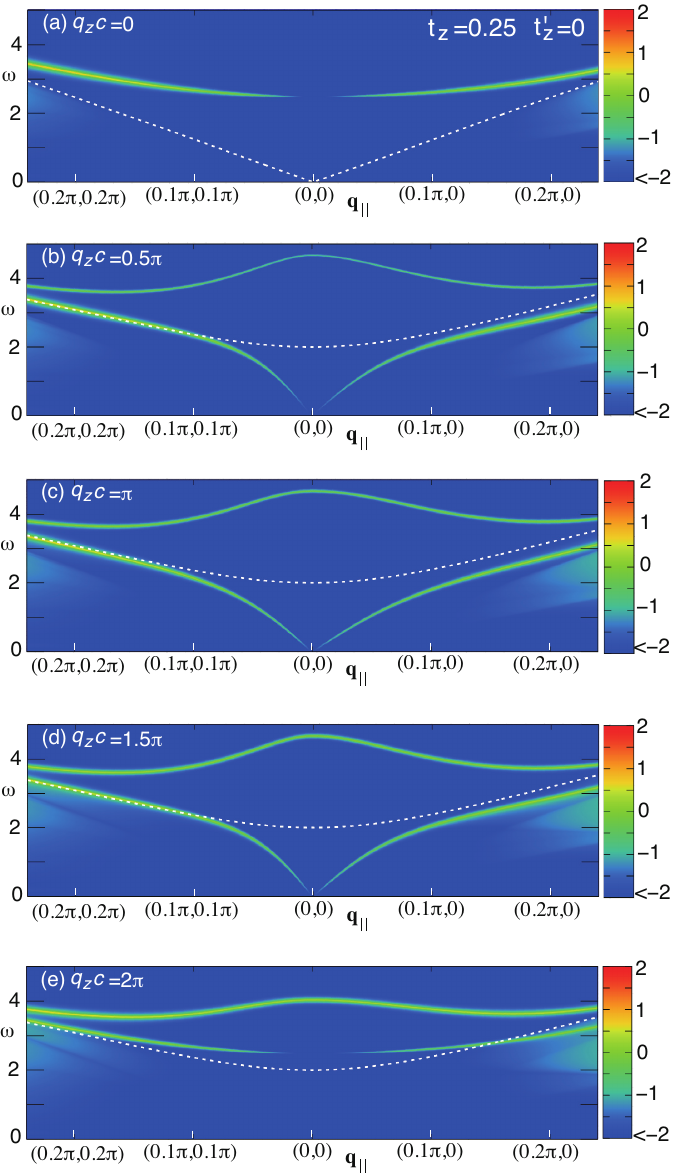}
\caption{Intensity maps of $\log_{10} | {\rm Im}\kappa(\vq, \omega)|$  for a sequence of $q_{z}c$ around a region of $\vq_{\parallel}=(0,0)$: (a) $q_{z}c=0$, (b) $q_{z}c=0.5\pi$, (c) $q_{z}c=\pi$, (d) $q_{z}c=1.5\pi$, and (e) $q_{z}c=2\pi$. The white dotted curve denotes the upper boundary of the particle-hole continuum. It becomes zero at $\vq_{\parallel}=(0,0)$ and $q_{z}c=0$ even for a finite $t_{z}=0.25$. To keep an appropriate contrast, the same color is used below -2. 
}
\label{kqw-tz025}
\end{figure}
%%%%%%%%%%%%%%%%%%%%%%%%%%%%%%%%%%%%%%%%%%%%%%%%%

We have performed the same calculations at $t_{z}=0$ and checked that results at $t_{z}=0$ hardly change from those at $t_{z}=0.01$ except that the $\omega_{+}$ ($\omega_{-}$) mode also becomes gapless at $q_{z}c \ne 2n\pi$ [$q_{z}c = 2n\pi (n\ne 0)$].

\subsubsection{A large value of $t_{z}$} 

%%%%%%%%%%%%%%%%%%%%% FIG. 8%%%%%%%%%%%%%%%%%%%%%%%%
\begin{figure}[t]
\centering
\includegraphics[width=8cm]{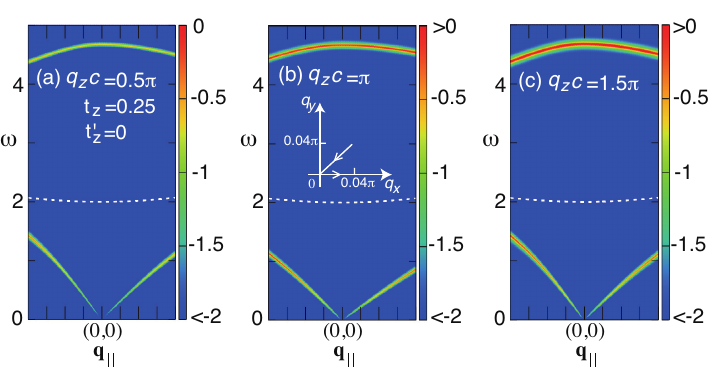}
\caption{Intensity maps of $\log_{10} | {\rm Im}\kappa(\vq, \omega)|$ in the vicinity of $\vq_{\parallel}=(0,0)$ along $(0.04\pi, 0.04\pi)$-$(0,0)$-$(0.04\pi, 0)$ [see the inset in (b)] for (a) $q_{z}c=0.5\pi$, (b) $q_{z}c=\pi$, and (c) $q_{z}c=1.5\pi$. It is the $\omega_{+}$ mode that becomes gapless and the $\omega_{-}$ mode is gapped for a large $t_{z}$.  The white dotted curve denotes the upper boundary of the particle-hole continuum. To keep an appropriate contrast, the same color is used below -2 and above 0. 
}
\label{kqw-tz025detail}
\end{figure}
%%%%%%%%%%%%%%%%%%%%%%%%%%%%%%%%%%%%%%%%%%%%%%%%%

We also study a large value of $t_{z}$ for completeness. Maps of the spectral function are shown in \fig{kqw-tz025} for a sequence of $q_{z}c$. At $q_{z}c=0$ [\fig{kqw-tz025}(a)], only one mode is realized as in the case of Figs.~\ref{kqw-00}(a) and \ref{kqw-tz001}(a). The spectral intensity at $\vq_{\parallel}=(0,0)$ is zero, indicating that it should be in-phase charge fluctuations between the two layers inside the unit cell because of the charge conservation. That is, it should be the $\omega_{+}$ mode. At $q_{z}c=0.5\pi$, there appear two modes as shown in \fig{kqw-tz025}(b). While this  feature is the same as that in \fig{kqw-00}(b), the upper mode forms a convex shape around $\vq_{\parallel}=(0,0)$ for $t_{z}=0.25$ whereas the other mode is gapless as clarified in \fig{kqw-tz025detail}(a). A similar feature is observed for other $q_{z}c$ in Figs.~\ref{kqw-tz025}(c)(d) and \ref{kqw-tz025detail}(b)(c). It is remarkable that this low-energy mode is sharply defined even inside the continuum. At $q_{z}c=2\pi$ [\fig{kqw-tz025}(e)], there appear two gapped modes. In contrast to the case of $t_{z}=0.1$ [\fig{kqw-00}(e)], it is the lower-energy  mode that is similar to the mode at $q_{z}c=0$ [\fig{kqw-tz025}(a)] and the spectral weight at $\vq_{\parallel}=(0,0)$ vanishes. This indicates that the mode with lower energy should  be the $\omega_{+}$ mode. Because of the continuity of the mode as a function of $q_{z}c$ (see also Sec.~III~A~6), it is natural to conclude that the $\omega_{+}$ mode becomes gapless in $q_{z}c\ne 2 n \pi$  in Figs.~\ref{kqw-tz025}(b)--(d) and \ref{kqw-tz025detail}(a)--(c) and the $\omega_{-}$ mode is always gapped independent of $q_{z}c$, except it vanishes at $q_{z}c=0$ [\fig{kqw-tz025}(a)]. Therefore, it is implied that the energy hierarchy of the $\omega_{\pm}$ modes is interchanged for a large $t_{z}$.

\subsubsection{$t_{z}$ dependence of the gap at $\vq_{\parallel} \approx (0,0)$}
The results for $t_{z}=0.01, 0.1, 0.25$ (Figs.~\ref{kqw-00} -- \ref{kqw-tz025detail}) imply that the gap of plasmon modes at $\vq_{\parallel} \approx (0,0)$ depends on $t_{z}$ in a nontrivial way. We here clarify its $t_{z}$ dependence.  

%%%%%%%%%%%%%%%%%%%%% FIG.  9 %%%%%%%%%%%%%%%%%%%%%%%%
\begin{figure}[th]
\centering
\includegraphics[width=7cm]{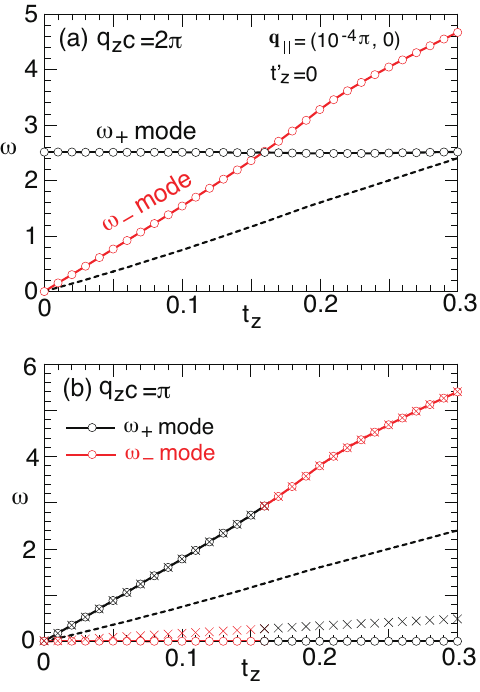}
\caption{$t_{z}$ dependence of the gap of the $\omega_{\pm}$ modes at $\vq_{\parallel}=(10^{-4}\pi, 0)$ for (a) $q_{z}c=2\pi$ and (b) $q_{z}c=\pi$. The dashed line is the upper boundary of the continuum spectrum. In (a), the $\omega_{\pm}$ modes are determined from the minimum position of \eq{detk}. The point $\omega_{-}=0$ at $t_{z}=0$ is introduced by hand for technical reasons.  In (b), crosses denote the minimum positions of \eq{detk}. They deviate from the $\omega_{\pm}$-mode energy at low energy because of the mixture of the particle-hole continuum. Note that the energy of the $\omega_{\pm}$ modes is interchanged across $t_{z} \approx 0.16$ because of the crossing of these modes at $q_{z}c=2n\pi$ as shown in (a). 
}
\label{gap-tz}
\end{figure}
%%%%%%%%%%%%%%%%%%%%%%%%%%%%%%%%%%%%%%%%%%%%%%%%

Figure~\ref{gap-tz}(a) is  representative results for $q_{z}c = 2n \pi$ with $n$ being integer---note that only the $\omega_{+}$ mode is present at $q_{z}c=0$, as already shown in Figs.~\ref{kqw-00}(a), \ref{kqw-tz001}(a), and \ref{kqw-tz025}(a). Since the $\omega_{\pm}$ modes are realized above the particle-hole continuum, we plot the minimum position of \eq{detk} in \fig{gap-tz}(a) to achieve stable computation. The energy of the $\omega_{+}$ mode is essentially independent of $t_{z}$ whereas the energy of the $\omega_{-}$ mode increases monotonically with $t_{z}$.  Interestingly for a large $t_{z} (\geq 0.16)$, the $\omega_{-}$ mode has energy higher than the $\omega_{+}$ mode. 
 
On the other hand, \fig{gap-tz}(b) is a representative result for $q_{z}c \ne 2 n \pi$, showing very different behavior to that at $q_{z}c =2 n \pi$. The energy of the $\omega_{+}$ mode increases monotonically with increasing $t_{z}$. Furthermore, the $\omega_{+}$ mode follows perfectly the minimum position of \eq{detk}  since it is always realized above the continuum. The energy of the $\omega_{-}$ mode is zero independent of $t_{z}$ as already indicated in Figs.~\ref{kqw-000} and \ref{kqw-tz001detail}. Its energy is deviated 
from the minimum position of \eq{detk} as increasing  $t_{z}$. This is because in contrast to the $\omega_{+}$ mode, the $\omega_{-}$ mode is realized inside the particle-hole continuum as shown by the dashed line in \fig{gap-tz}(b), when $\vq_{\parallel}$ is close to $(0,0)$. These descriptions hold up to $t_{z} \approx 0.15$. As seen in \fig{gap-tz}(a), the $\omega_{+}$ mode becomes lower than the $\omega_{-}$ mode in $t_{z} \geq 0.16$ at $q_{z}c=2n\pi$. Consequently, it is the $\omega_{+}$ mode that exhibits a gapless mode and the $\omega_{-}$ mode is gapped in $t_{z}\geq 0.16$ for $q_{z}c \ne 2 n \pi$. Therefore, by crossing $t_{z} \approx 0.16$, the energy hierarchy of the $\omega_{\pm}$ modes is interchanged as already implied  in \fig{kqw-tz025}.

\subsubsection{$q_{z}c$ dependence}

%%%%%%%%%%%%%%%%%%%%% FIG. 10  %%%%%%%%%%%%%%%%%%%%%%%%
\begin{figure}[b]
\centering
\includegraphics[width=16cm]{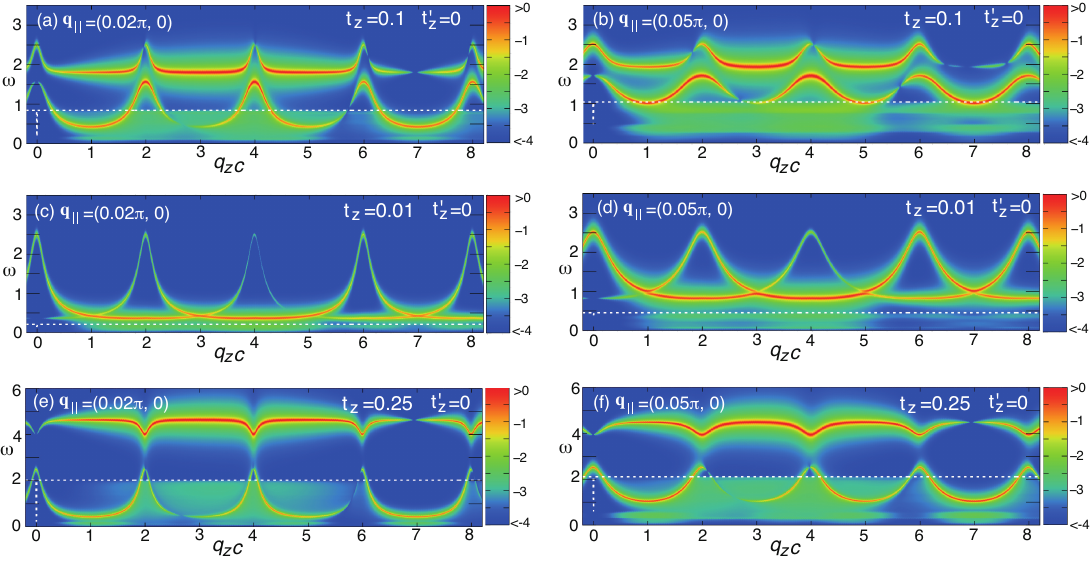}
\caption{$q_{z}c$ dependence of the intensity map of $\log_{10} | {\rm Im}\kappa(\vq, \omega)|$ at (a) $\vq_{\parallel}=(0.02\pi, 0)$ and (b) $\vq_{\parallel}=(0.05\pi, 0)$ for $t_{z}=0.1$. The corresponding results of $t_{z}=0.01$ and $0.25$ are shown in (c)(d) and (e)(f), respectively. To get a reasonable contrast, the same color is used above 0 and below -4. The white dotted line is the upper boundary of the continuum spectrum and exhibits a spike at $q_{z}c=0$ because of the vanishing of the $\omega_{-}$ mode there. 
}
\label{kqw-qzc}
\end{figure}
%%%%%%%%%%%%%%%%%%%%%%%%%%%%%%%%%%%%%%%%%%%%%%%%%
So far we have chosen certain points of $q_{z}c$. Here we clarify the $q_{z}c$ dependence of the $\omega_{\pm}$ modes. Figure~\ref{kqw-qzc}(a) shows results at $\vq_{\parallel}=(0.02\pi, 0)$ for $t_{z}=0.1$. A white dotted line is the upper boundary of the particle-hole continuum. The $\omega_{+}$ mode is always realized above the continuum. A peak at $q_{z}c = 2n\pi$ corresponds to the optical plasmon. The energy of the $\omega_{+}$ mode is quickly decreased as going away from $q_{z}c = 2n\pi$ and becomes essentially $q_{z}c$ independent until the next peak at $q_{z}c = 2(n+1) \pi$. The $\omega_{-}$ mode does not exist at $q_{z}c=0$, which is evident from the zero intensity there in \fig{kqw-qzc}(a). However, as going away from $q_{z}c=0$, the $\omega_{-}$ mode immediately gains the intensity and starts to show a dispersion between $\omega \approx 0.4$--$1.6$ with a peak at $q_{z}c = 2n\pi (n \ne 0)$. Interestingly this dispersive feature occurs crossing the upper boundary of the continuum. 

Figure~\ref{kqw-qzc}(b) is the same plot but  for $\vq_{\parallel}=(0.05\pi, 0)$. As we have seen in \fig{kqw-000}, the $\omega_{+}$ mode has a weak $\vq_{\parallel}$ dependence around  $\vq_{\parallel}=(0,0)$ and thus the result is similar to that in \fig{kqw-qzc}(a). On the other hand, the $\omega_{-}$ mode exhibits a cosinelike dispersion along the $q_{z}c$ direction  and is located above the continuum. 

In both Figs.~\ref{kqw-qzc}(a) and (b), the intensity of the $\omega_{\pm}$ modes has a characteristic $q_{z}c$ dependence. In particular, the $\omega_{+}$ mode loses intensity around $q_{z}c=7\pi$ and both $\omega_{\pm}$ modes have a kind of ``nodes'' at specific $q_{z}c$ values. These positions depend on the choices of parameters (except for the vanishing intensity of the $\omega_{-}$ mode at $q_{z}c=0$) and originate from the suppression of the numerator of Im$\kappa(\vq, \omega)$ in \eq{kqw1}. In other words, the absence of $2 \pi$ periodicity in \fig{kqw-qzc} originates from the numerator of \eq{kqw1}, namely, the factor of $\cos^{2}\frac{q_{z}d}{2}$, $\sin^{2}\frac{q_{z}d}{2}$, and $\frac{V^{'}(\vq) + V^{'}(-\vq)}{2}$. In fact, it is easily checked that the denominator of Im$\kappa(\vq, \omega)$ [\eq{kqw1det}], i.e., \eq{detk} retains $2\pi$ periodicity along the $q_{z}c$ direction; see Figs.~\ref{kqw-tzp-qzc}(c) and (d). 

The $q_{z}c$ dependence of the $\omega_{\pm}$ modes gives a different impression for a smaller $t_{z}$. We present corresponding results for $t_{z}=0.01$ in Figs.~\ref{kqw-qzc}(c) and (d). While the intensity of the $\omega_{-}$ mode vanishes at $q_{z}c=0$ as we have already explained, the $\omega_{-}$ mode has a much weaker dispersion---almost flat---than that at $t_{z}=0.1$. In particular, the $\omega_{-}$ mode does not exhibit a peak even at $q_{z}c = 2n\pi (n\ne 0)$, in a sharp contrast to the case of $t_{z}=0.1$. That is, the $\omega_{-}$ mode tends to have a weaker dispersive feature with decreasing $t_{z}$. In contrast, the $\omega_{+}$ mode has a much stronger dispersion between $\omega \approx 0.5$--$2.5$ in \fig{kqw-qzc}(c) and $\omega \approx 1$--$2.5$ in \fig{kqw-qzc}(d), compared with that for $t_{z}=0.1$ shown in Figs.~\ref{kqw-qzc}(a) and (b), respectively, even though $t_{z}$ is much smaller. As we have already seen in Figs.~\ref{kqw-tz001}(c) and \ref{kqw-tz001detail}(b), the $\omega_{\pm}$ modes almost overlap with each other at $q_{z}c=\pi + 2n \pi$. 

For a large $t_{z}$, as we have shown in \fig{gap-tz}, the $\omega_{-}$ mode has higher energy than the $\omega_{+}$ mode. This can also be confirmed by zero spectral weight at $q_{z}c=0$ in the higher-energy mode in Figs.~\ref{kqw-qzc}(e)(f)---a characteristic feature of the $\omega_{-}$ mode as we have already discussed in the context of \fig{kqw-00}(a). The $\omega_{-}$ mode shows the $q_{z}c$ dependence very different from Figs.~\ref{kqw-qzc}(a)--(d): the $\omega_{-}$ mode has a dip at $q_{z}c = 2 n \pi$, not a peak there, in Figs.~\ref{kqw-qzc}(e)(f), and an almost flat feature away from $q_{z}c = 2n \pi$. In particular, the dip of the $\omega_{-}$ mode touches with a peak of the $\omega_{+}$ mode at $q_{z}c = 2n\pi$ at $t_{z}  \approx 0.16$, where the energy of the $\omega_{\pm}$ mode is interchanged.  In contrast to the $\omega_{-}$ mode, the lower-energy mode, namely $\omega_{+}$ mode, has a large dispersive feature at $\vq_{\parallel}=(0.02\pi,0)$ between $\omega \approx 0.5$--$2.5$ and is sharply defined even though the $\omega_{+}$ mode is realized inside the continuum especially for $q_{z}c \ne 2n\pi$. This is simply due to low spectral weight of the continuum especially around $\vq_{\parallel} = (0,0)$.  At $\vq_{\parallel} = (0.05\pi, 0)$ shown in \fig{kqw-qzc}(f), the $\omega_{+}$ mode is similar to that at $\vq_{\parallel} = (0.02\pi,0)$ [\fig{kqw-qzc}(e)], but becomes less dispersive. 

One would expect a larger $q_{z}c$ dispersion for a larger $t_{z}$ because the system is closer to be three dimensional. However, Figs.~\ref{kqw-qzc}(a)(c)(e) do not support such a native expectation. The $\omega_{+}$ mode has a large dispersion in $0.5 \lesssim \omega \lesssim 2.5$ at $t_{z}=0.01$ and 0.25, but a smaller dispersion in $2 \lesssim \omega \lesssim 2.5$ at $t_{z}=0.1$. The $\omega_{-}$ mode, on the other hand,  it is almost dispersionless at $t_{z}=0.01$ and has a larger dispersion in $0.5 \lesssim \omega \lesssim 1.5$ at $t_{z}=0.1$, but an almost flat dispersion in $4 \lesssim \omega \lesssim 4.5$ at $t_{z}=0.25$. 

\subsection{Charge excitation spectra with \boldmath{$t_{z}^{'} \ne 0$}}
So far we have focused on the case of $t_{z}^{'}=0$, no interbilayer hopping, although electrons in all layers interact with each other via the LRC. Here we first focus on the case of $t_{z}=0.1$ and study the effect of $t_{z}^{'}$ by assuming $t_{z}^{'}=t_{z}/2$. Since results may become different for a large $t_{z}$ as seen in Figs.~\ref{kqw-tz025} and \ref{kqw-tz025detail}, such results shall be presented in the last section~$4$. 

%%%%%%%%%%%%%%%%%%%%% FIG.  11 %%%%%%%%%%%%%%%%%%%%%%%%
\begin{figure}[t]
\centering
\includegraphics[width=8cm]{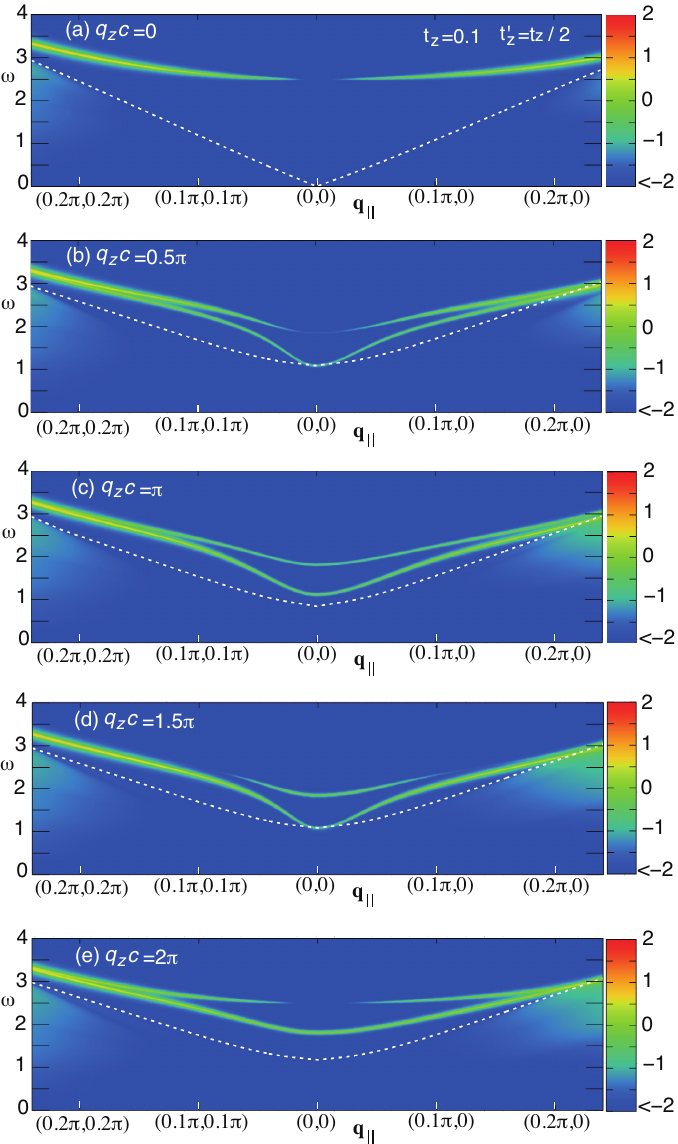}
\caption{Intensity map of $\log_{10} | {\rm Im}\kappa(\vq, \omega)|$ for a sequence of $q_{z}c$ around a region of $\vq_{\parallel}=(0,0)$: (a) $q_{z}c=0$, (b) $q_{z}c=0.5\pi$, (c) $q_{z}c=\pi$, (d) $q_{z}c=1.5\pi$, and (e) $q_{z}c=2\pi$. The interbilayer hopping $t_{z}^{'}$ is introduced as $t_{z}^{'}=t_{z}/2=0.05$. The white dotted curve denotes the upper boundary of the particle-hole continuum. There is no continuum spectrum at $\vq_{\parallel}=(0,0)$ and $q_{z}c=0$ even for a finite $t_{z}=0.1$ and $t_{z}^{'}=0.05$. To keep an appropriate contrast, the same color is used below -2. 
}
\label{kqw-tzp}
\end{figure}
%%%%%%%%%%%%%%%%%%%%%%%%%%%%%%%%%%%%%%%%%%%%%%%%%

%%%%%%%%%%%%%%%%%%%%% FIG.  12 %%%%%%%%%%%%%%%%%%%%%%%%
\begin{figure}[ht]
\centering
\includegraphics[width=8cm]{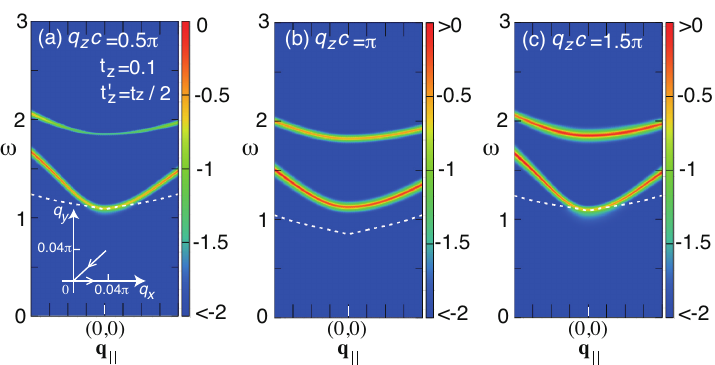}
\caption{Intensity maps of $\log_{10} | {\rm Im}\kappa(\vq, \omega)|$  in the vicinity of $\vq_{\parallel}=(0,0)$ along $(0.04\pi, 0.04\pi)$-$(0, 0)$-$(0.04\pi, 0)$ [see the inset in (a)] for various choices of $q_{z}c$: (a) $q_{z}c=0.5\pi$, (b) $q_{z}c=\pi$, and (c) $q_{z}c=1.5\pi$. The interbilayer hopping $t_{z}^{'}$ is introduced as $t_{z}^{'}=t_{z}/2=0.05$. The white dotted curve denotes the upper boundary of the particle-hole continuum. To keep an appropriate contrast, the same color is used above 0 and below -2. 
}
\label{kqw-tzpdetail}
\end{figure}
%%%%%%%%%%%%%%%%%%%%%%%%%%%%%%%%%%%%%%%%%%%%%%%%%

%%%%%%%%%%%%%%%%%%%%% FIG. 13  %%%%%%%%%%%%%%%%%%%%%%%%
\begin{figure}[t]
\centering
\includegraphics[width=7cm]{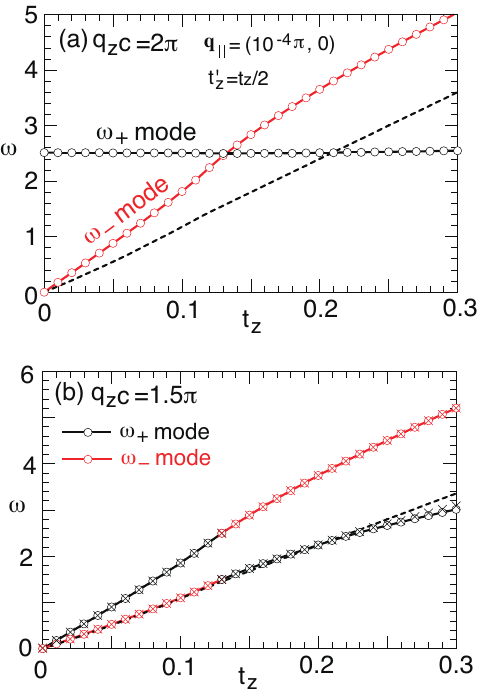}
\caption{$t_{z}$ dependence of the gap of the $\omega_{\pm}$ modes at $\vq_{\parallel}=(10^{-4}\pi, 0)$ at (a) $q_{z}c=2\pi$ and (b) $q_{z}c=1.5\pi$. The dashed line is the upper boundary of the continuum spectrum. In (a), the $\omega_{\pm}$ modes are determined from the minimum position of \eq{detk}. Since the $\omega_{+}$ mode is located below the continuum in $t_{z}\gtrsim 0.2$, these energies are estimated approximately. The point $\omega_{-}=0$ at $t_{z}=0$ is introduced by hand for technical reasons.  In (b), crosses denote the minimum position of \eq{detk}; only one minimum position is found at $t_{z}=0.01$. 
}
\label{gap-tzp-tz}
\end{figure}
%%%%%%%%%%%%%%%%%%%%%%%%%%%%%%%%%%%%%%%%%%%%%%%%%

\subsubsection{Maps of the spectral weight}
Results corresponding to Figs.~\ref{kqw-00}(a)--(e) are shown in Figs.~\ref{kqw-tzp}(a)--(e) in the same condition except for a finite $t_{z}^{'}=t_{z}/2=0.05$. At $q_{z}c=0$ [\fig{kqw-tzp}(a)], only the $\omega_{+}$ mode is realized and the spectrum is essentially the same as \fig{kqw-00}(a), that is, the effect of $t_{z}^{'}$ is negligible at $q_{z}c=0$. In fact, mathematically one can check that we obtain the same expression as \eq{kqw1qzc0-2} at $q_{z}c=0$. The negligible effect of $t_{z}^{'}$ also applies to the result at $q_{z}c=2\pi$ [\fig{kqw-tzp}(e)], except that the $\omega_{-}$ mode has a slightly larger energy because of the inclusion of  $t_{z}^{'}$. The major effect of $t_{z}^{'}$ is recognized at $q_{z}c \ne 2n\pi$. Whereas the $\omega_{+}$ mode exhibits a dispersion very similar to Figs.~\ref{kqw-00}(b)--(d), the $\omega_{-}$ mode is gapped at $\vq_{\parallel}=(0,0)$ by a finite $t_{z}^{'}$. Consequently we obtain two gapped modes at any $q_{z}c (\ne 0)$. 

To emphasize this qualitative change, we also show the zoom of the spectrum around $\vq_{\parallel} = (0,0)$ in Figs.~\ref{kqw-tzpdetail}(a)--(c). These results clearly demonstrate that the spectrum of the $\omega_{-}$ mode becomes gapped by including $t_{z}^{'}$ especially in the vicinity of $\vq_{\parallel}=(0,0)$, in sharp contrast to Figs.~\ref{kqw-000}(a)--(c).  

\subsubsection{$t_{z}$ dependence of the gap at $\vq_{\parallel} \approx (0,0)$} 
Next we clarify how the gap of the $\omega_{\pm}$ modes at $\vq_{\parallel}=(0,0)$ evolves with $t_{z}$ and $t_{z}^{'} (=t_{z}/2)$. Figure~\ref{gap-tzp-tz}(a) shows a representative result at $q_{z}c = 2n\pi$ although the $\omega_{-}$ mode vanishes at $q_{z}c=0$. Here, the gap energy of the $\omega_{\pm}$ modes is calculated from the minimum position of \eq{detk} to achieve stable computation. This is justified because the $\omega_{\pm}$ modes are realized above the continuum at least up to $t_{z}=0.2$. In $t_{z} > 0.2$, the actual energy of the $\omega_{+}$ mode may slightly deviate from the result in \fig{gap-tzp-tz}(a). The $t_{z}$ dependence of the $\omega_{+}$ mode can be negligible even in the presence of $t_{z}^{'}$ and the result is essentially the same as \fig{gap-tz}(a). The $\omega_{-}$ mode also exhibits a similar $t_{z}$ dependence to  \fig{gap-tz}(a) and the effect of $t_{z}^{'}$ is not crucial at $q_{z}c=2n\pi$. 

However, the presence of $t_{z}^{'}$ has a big effect for $q_{z}c \ne 2n\pi$ and the gapless mode disappears upon a finite $t_{z}$ and $t_{z}^{'}=t_{z}/2$, in sharp contrast to \fig{gap-tz}(b). Figure~\ref{gap-tzp-tz}(b) shows results at $q_{z}c=1.5\pi$. Both $\omega_{\pm}$ modes have higher energy with increasing $t_{z}$. A comparison with \fig{gap-tz}(b) confirms that the gap of the $\omega_{-}$ ($\omega_{+}$) mode in $t_{z} \leq 0.13$ ($\geq 0.13$)  originates from a finite $t_{z}^{'}$. 

In \fig{gap-tzp-tz}(b), the energy of the $\omega_{-}$ mode is located close to the upper boundary of the particle-hole continuum in $t_{z} \leq 0.13$. In $t_{z} \geq 0.13$, however, the $\omega_{+}$ mode has lower energy than the $\omega_{-}$ mode and is located close to the upper boundary of the continuum. In both regions, their energy position is well captured by the minimum position of \eq{detk} shown by crosses. If we choose $q_{z}c=\pi$ and would make the same plot as \fig{gap-tzp-tz}(b), the plasmon modes would be realized above the continuum even for a large $t_{z}$, which may be readily inferred from Figs.~\ref{kqw-tzp}(c) and \ref{kqw-tzpdetail}(b).

\subsubsection{$q_{z}c$ dependence} 
%%%%%%%%%%%%%%%%%%%%% FIG. 14  %%%%%%%%%%%%%%%%%%%%%%%%
\begin{figure}[t]
\centering
\includegraphics[width=16cm]{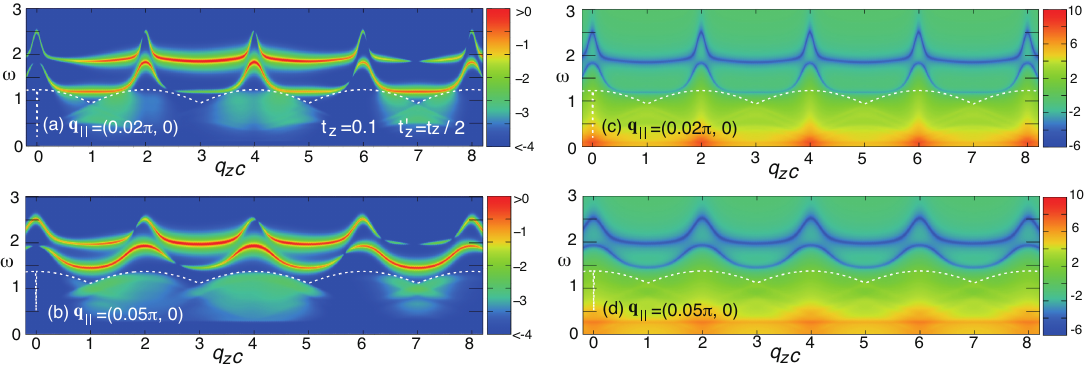}
\caption{$q_{z}c$ dependence of the intensity map of $\log_{10} | {\rm Im}\kappa(\vq, \omega)|$ (left panels) and $\log_{10}|{\mathfrak{det} }|^{2}$ given in \eq{det2} (right panels) at $\vq_{\parallel}=(0.02\pi, 0)$ (a) and (c), and $\vq_{\parallel}=(0.05\pi, 0)$ (b) and (d) for $t_{z}^{'}=t_{z}/2=0.05$. To get a reasonable contrast, the same color is used above 0 and below -4 in the left panels. The white dotted curve is the upper boundary of the continuum spectrum and exhibits a spike at $q_{z}c=0$ where the $\omega_{-}$ mode vanishes. 
}
\label{kqw-tzp-qzc}
\end{figure}
%%%%%%%%%%%%%%%%%%%%%%%%%%%%%%%%%%%%%%%%%%%%%%%%%

The $q_{z}c$ dependence of the $\omega_{\pm}$ modes is shown in  Figs.~\ref{kqw-tzp-qzc}(a) and (b) at $\vq_{\parallel}=(0.02\pi, 0)$ and $(0.05\pi, 0)$, respectively. By comparing with Figs.~\ref{kqw-qzc}(a)(b), essentially the same results are obtained except that the energy of the $\omega_{-}$ mode is pushed up to higher energy by a finite $t_{z}^{'}$. While the intensity of the spectrum does not have $2\pi$ periodicity as discussed already in the context of \fig{kqw-qzc}, it is interesting that the spectrum is strongly suppressed at almost the same positions of $q_{z}c$ shown in Figs.~\ref{kqw-qzc}(a)(b). That is, the effect of $t_{z}^{'}$ is weak for the intensity dependence of the $\omega_{\pm}$ modes. 

We can check analytically that the denominator of $\kappa(\vq, \omega)$ in \eq{kqw2}, i.e., \eq{detk} has $2\pi$ periodicity with respect to $q_{z}c$ and we show such a property in the right panels in \fig{kqw-tzp-qzc}, where its square of the modulus is presented. Therefore, the non-$2\pi$ periodicity of the intensity of the Im$\kappa(\vq, \omega)$ originates from the numerator, namely, $\kappa^{0}_{\cos}$ and $V^{'}_{+}(\vq)$ in \eq{kqw2}.

\subsubsection{Spectra for a large $t_{z}$}  

As we have seen in Sec.~III~A, the energy hierarchy of the $\omega_{\pm}$ modes is interchanged for a large $t_{z}$. This also happens in the presence of $t_{z}^{'}$ as already shown in \fig{gap-tzp-tz}. The presence of $t_{z}^{'}$ reduces the crossing value of $t_{z}$ down to $t_{z}=0.13$ when $t_{z}^{'}=t_{z}/2$. This value of $t_{z}$ may not be regarded so large and can be realistic for some bilayer materials. Hence, it should be useful to present how the spectrum looks like for a large $t_{z}$. To get a clear separation between the $\omega_{\pm}$ modes, we here choose $t_{z}=0.25$. 

%%%%%%%%%%%%%%%%%%%%% FIG. 15  %%%%%%%%%%%%%%%%%%%%%%%%
\begin{figure}[th]
\centering
\includegraphics[width=8cm]{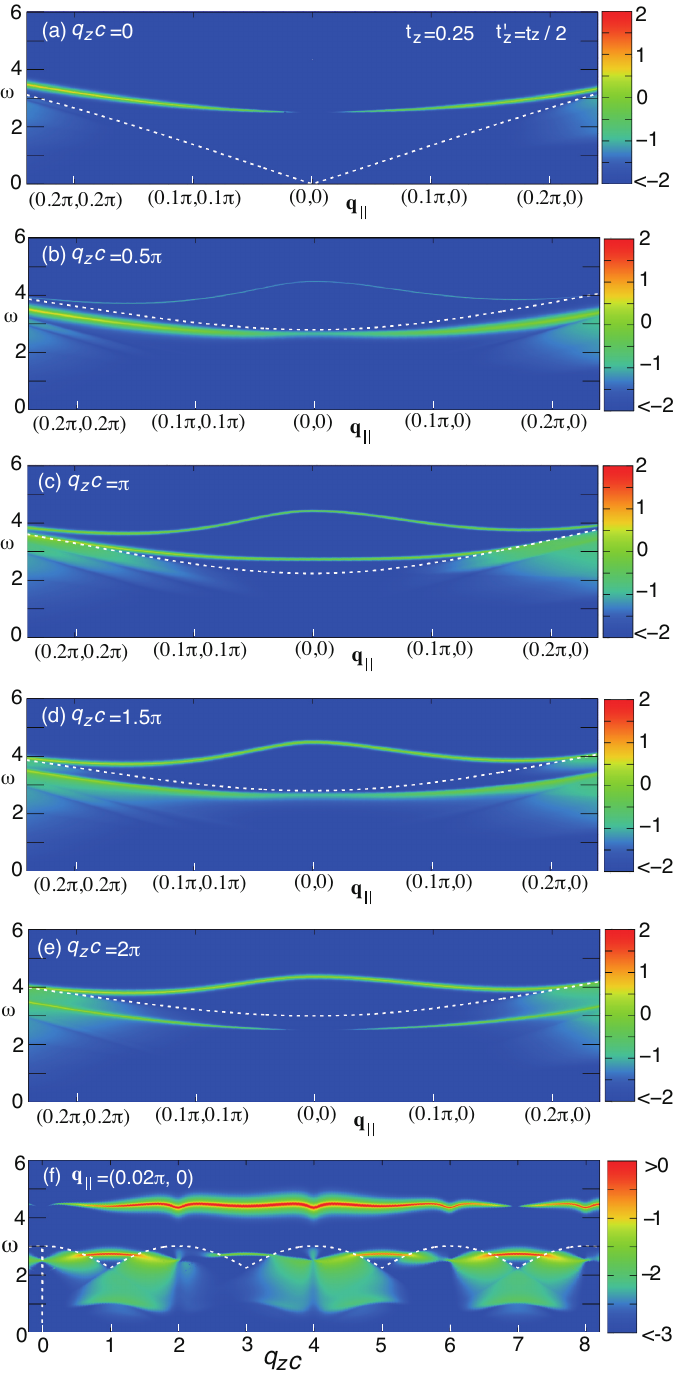}
\caption{Intensity map of $\log_{10} | {\rm Im}\kappa(\vq, \omega)|$ in the presence of a large $t_{z}=0.25$ and $t_{z}^{'}=t_{z}/2$. (a)--(e) $\vq_{\parallel}$ dependence for a sequence of $q_{z}c$ around a region of $\vq_{\parallel}=(0,0)$: (a) $q_{z}c=0$, (b) $q_{z}c=0.5\pi$, (c) $q_{z}c=\pi$, (d) $q_{z}c=1.5\pi$, and (e) $q_{z}c=2\pi$. (f) $q_{z}c$ dependence of the intensity map of $\log_{10} | {\rm Im}\kappa(\vq, \omega)|$ at $\vq_{\parallel}=(0.02\pi, 0)$.  The white dotted curve denotes the upper boundary of the particle-hole continuum---there is no continuum spectrum at $\vq_{\parallel}=(0,0)$ and $q_{z}c=0$ even for a finite $t_{z}$ and $t_{z}^{'}$ in (a) and there is a spike at $q_{z}c=0$ in (f) because of the disappearance of the $\omega_{-}$ mode there. To keep an appropriate contrast, the same color is used below -2 in (a)--(e), and above 0 and below -3 in (f). 
}
\label{kqw-tzp025}
\end{figure}
%%%%%%%%%%%%%%%%%%%%%%%%%%%%%%%%%%%%%%%%%%%%%%%%%

$\vq$-$\omega$ maps are shown in Figs.~\ref{kqw-tzp025}(a)--(e) for a sequence of $q_{z}c$. At $q_{z}c=0$  [\fig{kqw-tzp025}(a)], we have only the $\omega_{+}$ mode and the spectral weight at $\vq_{\parallel}=(0,0)$ vanishes due to the charge conservation as we already explained in Sec.~III~A~2; the upper boundary of the particle-hole excitations is zero at $\vq_{\parallel}=(0,0)$. With increasing $q_{z}c$, the particle-hole continuum gains the spectral weight even at $\vq_{\parallel}=(0,0)$ and the $\omega_{+}$ mode is realized just below its boundary. The $\omega_{-}$ mode has higher energy than the $\omega_{+}$ mode and is located above the continuum. However, the proximity to $q_{z}c=0$ yields tiny spectral weight and the $\omega_{-}$ mode is barely visible with a convex upward centered at $\vq_{\parallel} = (0,0)$ and $\omega \approx 4.5$ in \fig{kqw-tzp025}(b). This $\omega_{-}$ mode becomes more visible with increasing $q_{z}c$ as shown in Figs.~\ref{kqw-tzp025}(c)(d)(e) with a small $q_{z}c$ dependence. The $\omega_{+}$ mode has energy lower than the $\omega_{-}$ mode and is realized just above the continuum spectrum at $q_{z}c=\pi$ [\fig{kqw-tzp025}(c)], but enters into the continuum for larger $q_{z}c$ as seen in Figs.~\ref{kqw-tzp025}(d)(e). The reason why the $\omega_{+}$ mode is relatively sharp even inside the continuum lies in small spectral weight of the continuum around $\vq_{\parallel}=(0,0)$: it tends to be broadened around $\vq_{\parallel} = (0.2\pi, 0.2\pi)$ and $(0.2\pi,0)$, sufficiently away from $(0,0)$. 

In \fig{kqw-tzp025}(f), we present the $q_{z}c$ dependence of the spectral intensity at $\vq_{\parallel} = (0.02\pi, 0)$. It is interesting that both $\omega_{\pm}$ modes have a smaller dispersion in spite of a large $t_{z}$ and the system becomes more three dimensional by comparing with \fig{kqw-tzp-qzc}. The $\omega_{+}$ mode is sharply defined above the continuum around $q_{z}c=\pi + 2n\pi$ although its intensity depends on $q_{z}c$. Around $q_{z}c=2n\pi$, on the other hand, the $\omega_{+}$ mode is blurred substantially by the mixture of the continuum spectrum. The $\omega_{+}$ mode can be traced better in Figs.~\ref{kqw-tzp025}(a)--(e) rather than the $q_{z}c$ dependence in \fig{kqw-tzp025}(f). The $\omega_{-}$ mode vanishes at $q_{z}c=0$ and exhibits an almost flat dispersion as a function of $q_{z}c$ in \fig{kqw-tzp025}(f). In particular, as seen in Figs.~\ref{kqw-qzc}(e)(f), the $\omega_{-}$ mode exhibits a dip at $q_{z}c=2n\pi (n \neq 0)$. The $\omega_{-}$ mode is hardly visible at $q_{z}c=7\pi$, which is due to the band parameters, especially a specific value of $d/c$, and has been already observed in Figs.~\ref{kqw-qzc}(e)(f).

\section{Discussion}
\subsection{Comparison with RIXS data}
We are aware of RIXS data of plasmon excitations in Y-based cuprates \cite{bejas24}. These data were analyzed by employing the LRC obtained in the electron-gas model \cite{fetter74,griffin89}. Although such an analysis may not be justified in a strict sense for a system near half-filling such as in cuprates (electron-liquid systems), the authors expected to clarify whether the experimental data may be related to the $q^{-2}$ singularity of the LRC, the essential feature to realize plasmon excitations, namely, a  physics of the long-wavelength limit.  

\begin{figure}[tbh]
\centering
\includegraphics[width=8cm]{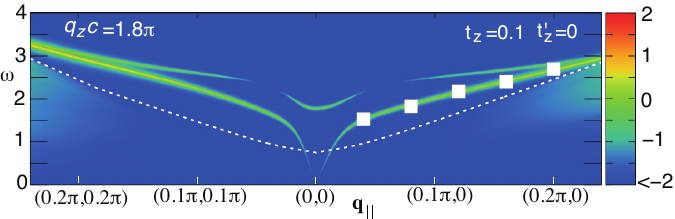}
\caption{Comparison with the plasmon energy (solid squares) reported in Y-based cuprate superconductors in Ref.~\cite{bejas24}; $t$ is assumed to be 233 meV. The experimental data are superimposed on the intensity map of $\log_{10} | {\rm Im}\kappa(\vq, \omega)|$ at $q_{z}c=1.8\pi$, $t_{z}=0.1$, and $t_{z}^{'}=0$. The white dotted curve denotes the upper boundary of the particle-hole continuum. To keep an appropriate contrast, the same color is used below -2. 
}
\label{hepting-fit}
\end{figure}

The present theory extends the Fetter model \cite{fetter74,griffin89} by deriving the LRC [Eqs.~(\ref{Vq}) and (\ref{Vqp})] in the bilayer lattice structure and making it applicable to any electron density. Hence it is interesting to confirm that the experimental data are indeed interpreted as plasmon excitations. We can find a parameter set to reproduce the experimental data as shown in \fig{hepting-fit}  when we choose $t=233$ meV. This value is smaller than the value obtained in {\it ab initio} calculations \cite{hybertsen90,andersen95,markiewicz05}. Our $t$ should be interpreted as an effective one and thus becomes smaller than the bare value, especially when making a comparison with experimental data \cite{nag24}. Our results partly share those given in Ref.~\cite{bejas24}, but important differences are revealed. 

%%%%%%%%%%%%%%%%%%%%% FIG. 16  %%%%%%%%%%%%%%%%%%%%%%%%
%%%%%%%%%%%%%%%%%%%%%%%%%%%%%%%%%%%%%%%%%%%%%%%%%

At $q_{z}c=1.8\pi$, the $\omega_{-}$ mode has higher intensity than the $\omega_{+}$ mode and the experimental data follow the $\omega_{-}$ mode. The same conclusion was obtained by studying the coupling to the superconducting phase to the electromagnetic fields \cite{sellati24}. In Ref.~\cite{bejas24}, $q_{z}c=0.2\pi$ was considered by assuming $2\pi$ periodicity with respect to $q_{z}c$ and it was concluded that the $\omega_{+}$ mode followed the experimental data.  However, as we have shown explicitly in \fig{kqw-qzc}, the plasmon intensity can have a strong $q_{z}c$ dependence. In particular, as we have emphasized in the context of Figs.~\ref{kqw-00}(a) and (e), $q_{z}c=0$ is a special point and the $\omega_{-}$ mode disappears there, but not $q_{z}c = 2n \pi (n \ne 0)$. The modulation of the peak intensity originates from the numerator of \eq{kqw1} via the factor $q_{z}d$, namely the geometry of the lattice structure. Nonetheless, we cannot deny a possibility that we could find another parameter set which allows the experimental data to be interpreted as the $\omega_{+}$ mode. In addition, one should be careful because the energy hierarchy of the $\omega_{\pm}$ modes is interchanged for $t_{z} \geq 0.16$ (0.13 for $t_{z}^{'} =t_{z}/2$). To clarify which mode the RIXS data indicate, $\omega_{+}$ mode or $\omega_{-}$ mode, further RIXS experiments are definitely necessary. 

A practical procedure to identify the $\omega_{\pm}$ mode in RIXS measurements would be the following. Choose $q_{z}c=2n\pi (n \ne 0)$ and $\vq_{\parallel}\approx (0,0)$. When RIXS detects collective charge excitations, they should be related with  the $\omega_{-}$ mode since the spectral weight of the $\omega_{+}$ mode is expected to be negligible there [see Figs.~\ref{kqw-00}(e) and \ref{kqw-tzp}(e)]. After that, $q_{z}c$ will be moved away from $2n\pi (n \ne 0)$ so that RIXS can trace the $\omega_{-}$ mode. At the same time, some additional signal is expected above or below the $\omega_{-}$ mode, depending on a value of $t_{z}$, which should be associated with the $\omega_{+}$ mode. 

To pursue a quantitative comparison with experiments, strong correlation effects should also be considered especially for cuprates and the obtained LRC on the bilayer lattice [Eqs.~(\ref{Vq}) and (\ref{Vqp})] should be incorporated into the framework of the so-called $t$-$J$-$V$ model \cite{greco16}, which is under way. 

\subsection{Comparison with the Fetter model} 
The Fetter model \cite{fetter74} was employed in Ref.~\cite{griffin89} to study plasmon excitations in bilayer cuprates. Since $t_{z}=0$ was taken in Ref.~\cite{griffin89}, we may compare our results at $t_{z}=0.01$, which shows nearly the same results as those at $t_{z}=0$. 

In Ref.~\cite{griffin89}, they can take the limit of $d \rightarrow 0$, yielding $V(\vq) = V^{'}(\vq)$. This is a specific property of the electron-gas model and the electron density becomes double in this limit. In the present theory, however, the model is defined in $0 < d \leq c/2$. Although one can invoke a small $d$, the limit of $d=0$ is not defined; instead, one should use a single-layer lattice model for $d=0$, not the present model. 

In Ref.~\cite{griffin89}, there is no $2 \pi$ periodicity along the $q_{x}$ and $q_{y}$ directions whereas the present theory describes such $2 \pi$ periodicity as it should do in crystals. In contrast,  the periodicity was assumed along the $q_{z}$ direction in Ref.~\cite{griffin89} and the value of $q_{z}$ is restricted in the first Brillouin zone. However, because of the multiple-point basis inside the unit cell along the $z$ direction, the spectral weight of charge excitations, i.e., Im$\kappa(\vq, \omega)$ does not have $2 \pi$ periodicity along the $q_{z}$ direction in the present model. A remarkable feature is that the denominator of the charge response function [Eqs.~(\ref{kqw1det}), (\ref{detk}), and (\ref{det2})] retains $2 \pi$ periodicity along the $q_{z}c$ direction even with a finite $t_{z}^{'}$ (see \fig{kqw-tzp-qzc}). This $2\pi$ periodicity is also present in Ref.~\cite{griffin89}. 

In the limit of $\vq \rightarrow {\bf 0}$, Ref.~\cite{griffin89} yields 
\be
V(\vq)=V^{'}(\vq) \propto \frac{1}{(q_{x}c)^{2}+(q_{y}c)^{2}+(q_{z}c)^{2}}\,,
\ee
namely, an isotropic Coulomb interaction. Note that the lattice constant $c$ appears behind $q_{x}$ and $q_{y}$ in their formalism. Instead, we obtain 
\be
V(\vq)=V^{'}(\vq) \propto \frac{1}{\alpha \left[ (q_{x}a)^{2} + (q_{y}a)^{2} \right] +(q_{z}c)^{2}}\,,
\ee
respecting the lattice anisotropy even in the limit of $\vq \rightarrow {\bf 0}$.

The $\omega_{-}$ mode was discussed to show essentially no $q_{z}$ dependence in Ref.~\cite{griffin89}. We have obtained similar results when $t_{z}$ is small; see Figs.~\ref{kqw-qzc}(c) and (d). However, as we have shown in Figs.~\ref{kqw-qzc} and \ref{kqw-tzp-qzc}, the $\omega_{-}$ mode exhibits a sizable $q_{z}$ dependence once a moderate $t_{z}$ is introduced; in particular, the $\omega_{-}$ mode has a peak at $q_{z}c=2n\pi$.

\subsection{Short-range interaction without \boldmath{$t_{z}'$}} 
Although we have considered the LRC, Eqs.~(\ref{HI}) and (\ref{RPA}) are valid for any functional form of $V(\vq)$ and $V^{'}(\vq)$. It may be informative to consider a short-range Coulomb interaction such as 
\bea
&&V(\vq)= 2 V_{xy}(\cos q_{x} + \cos q_{y}) \,, \\
&& V^{'}(\vq) = V_{z} {\rm e}^{-i q_{z} d} \,. 
\eea
We may assume no interbilayer interaction. The resulting susceptibility has then the same structure as that in the non-interaction case: 
\be
\kappa(\vq, \omega) = \cos^{2} \frac{q_{z}d}{2}  \kappa_{\rm even}(\vq,  \omega)+  \sin^{2} \frac{q_{z}d}{2} \kappa_{\rm odd}(\vq, \omega) \, ,
\label{short-eo}
\ee
where
\bea
&& \kappa_{\rm even}(\vq, \omega) = \frac{\kappa^{0}_{\rm even}(\vq, \omega)} {1-(V(\vq) + V_{z}) \kappa^{0}_{\rm even}(\vq, \omega) }
\label{short-e} \,, \\
&&  \kappa_{\rm odd}(\vq, \omega) = \frac{\kappa^{0}_{\rm odd}(\vq, \omega)} {1-(V(\vq) - V_{z}) \kappa^{0}_{\rm odd}(\vq, \omega) }
\label{short-o} \,.
\eea

It is straightforward to do similar calculations for short-range magnetic interactions such as $J(\vq)=2 J ( \cos q_{x} +  \cos q_{y})$  and $J^{'}(\vq) = J_{z} {\rm e}^{-i q_{z}d}$, yielding the dynamical magnetic susceptibility with the same functional forms as Eqs.~(\ref{short-eo}), (\ref{short-e}), and (\ref{short-o}). This describes the so-called bilayer modulation, which was actually observed by inelastic neutron scattering experiments for Y-based cuprates \cite{pailhes03,pailhes06}.

\subsection{Charge ordering tendencies in cuprates} 
As mentioned in the Introduction, besides plasmons, charge ordering tendencies \cite{ghiringhelli12,hashimoto14,peng16,chaix17,arpaia19,yu20,wslee21,lu22,arpaia23,da-silva-neto15,da-silva-neto16,da-silva-neto18} including the spin-charge stripe order \cite{tranquada95,tranquada21} were observed in cuprate superconductors. The present comprehensive analysis with the long-range Coulomb interaction cannot capture those ordering tendencies. This suggests at least two possibilities. One is that those phenomena are driven by higher-order electron correlations beyond the RPA. The other is that those are not usual charge excitations that we have studied here, but related to bond-charge excitations whose energy scale is controlled by the spin exchange interaction $J$ \cite{bejas17,zafur24}, indicating that the short-range magnetic interaction is also important. This idea indeed explained the data in electron-doped cuprates \cite{yamase15b,li17,yamase19}, but not in hole-doped cuprates  \cite{bejas12,allais14,meier14,wang14,atkinson15,yamakawa15,mishra15,zeyher18}. A possible reason may lie in a fact that the charge ordering tendency was observed inside the pseudogap phase in the hole-doped cuprates and the effect of the pseudogap should have been considered more seriously.

\section{Conclusions}
We have performed a comprehensive study of charge excitations in a bilayer system with the LRC and obtain important formulae and insights beyond the existing knowledge. There are two major results in the present work. 

The first one is a derivation of the LRC that fully respects the bilayer lattice structure and can be applicable to any electron density [Eqs.~(\ref{Vq}) and (\ref{Vqp})]. This is an extension of the Fetter model formulated 50 years ago \cite{fetter74} and the differences to the outcome from the Fetter model are clarified in Sec.~IV~B. We also provide a general formula of the charge susceptibility for the LRC [\eq{kqw2}] and compare it with a short-range Coulomb-interaction case (see Sec.~IV~C). While we have discussed the data of Y-based cuprates in Sec.~IV~A, the present theory is general and can be applicable to other bilayer materials.

The second one is to elucidate numerically the charge excitation spectrum for the bilayer system where the LRC is dominant; see Sec.~III. The spectral weight of plasmon dispersion loses $2 \pi$ periodicity along the $q_{z}c$ direction. However, it is remarkable to find that \eq{detk} retains $2 \pi$ periodicity even in the presence of the bilayer structure; we may indicate that a {\it fictitious} plasmon dispersion, not the spectral weight, has $2\pi$ periodicity in the $q_{z}c$ direction. As already pointed out in Ref.~\cite{griffin89}, the bilayer system hosts two modes: $\omega_{+}$ and $\omega_{-}$ mode. Although electrons in each layer interact with each other via the LRC, we may turn off the interbilayer hopping $t_{z}^{'}$ approximately for some materials. When $t_{z}$ is small, the $\omega_{+}$ mode has higher energy than the $\omega_{-}$ mode and is gapped whereas the $\omega_{-}$ mode is gapped only at $q_{z}c = 2 n \pi (n \ne 0)$ and exhibits a gapless mode for $q_{z}c \ne 2 n \pi$ at $\vq_{\parallel}=(0,0)$. Interestingly when $t_{z}$ becomes moderately large, the situation becomes vice versa: the $\omega_{-}$ mode has higher energy than the $\omega_{+}$ mode and is gapped whereas the $\omega_{+}$ mode is gapless for $q_{z}c \neq 2n\pi$ and gapped at $q_{z}c = 2n\pi$. The inclusion of $t_{z}^{'}$ yields a gap in both $\omega_{\pm}$ modes for $q_{z}c \neq 2n\pi$ whereas the modes at $q_{z}c = 2n\pi$ are  hardly affected by $t_{z}^{'}$. An unexpected feature is that when the system becomes more three dimensional by introducing a larger $t_{z}$, the $q_{z}c$ dispersion of the modes is not necessarily enhanced.  

The determination of the mode observed by RIXS is not straightforward. While we have found a parameter set that the recent RIXS data for Y-based cuprates \cite{bejas24} can be well explained by the $\omega_{-}$ mode, we cannot eliminate a possibility that other parameter sets could explain the data in terms of the $\omega_{+}$ mode. More detailed analyses are necessary both theoretically and experimentally to determine the mode observed in RIXS \cite{bejas24}, including a role of strong correlation effects beyond the present RPA. There is a subtlety that the energy hierarchy of the $\omega_{\pm}$ modes is interchanged by a moderately large, yet not unrealistic value of $t_{z}$. We have therefore proposed an idea to distinguish the $\omega_{\pm}$ modes in RIXS in Sec.~IV~A.

\acknowledgments
The author thanks P. Jakubczyk, G. Khaliullin,  W. Metzner, and M. Nieszporski for valuable discussions about how to formulate the LRC for the bilayer system, and M. Bejas and A. Fujimori for insightful discussions about plasmons in a bilayer system, and A. Greco and M. Hepting for constructive comments about the manuscript. The author is indebted to warm hospitality of Max-Planck-Institute for Solid State Research. He was supported by JSPS KAKENHI Grant No.~JP20H01856 and World Premier International  Research Center Initiative (WPI), MEXT, Japan.

\appendix

\section{Limit of \boldmath{$d \rightarrow c/2$}} 
The general expression of Eqs.~(\ref{Vq}), (\ref{Vqp}), and $\kappa(\vq, \omega)$ in \eq{kqw2} should be reduced to simpler expressions in the limit of $d \rightarrow c/2$. This may be a stringent test of the formalism given in the main text and thus we shall demonstrate that. 

In the case of $c=2d$ as well as $t_{z}=t_{z}^{'}$, we obtain in \eq{phi2}
\be
\varphi_{\vk \, \vk+\vq} = \left\{ \begin{aligned} 
&0 \quad {\rm for} \; \cos(k_{z}d+q_{z}d) \geq 0 \\
& \pi \quad {\rm for} \;\cos(k_{z}d+q_{z}d) < 0
\end{aligned}  
\right.
\ee
which lead to $\kappa^{0}_{\rm sin} =0$ in \eq{sink}. We then obtain the following simple expression: 
\be
\kappa^{0}_{11} + \kappa^{0}_{\cos} = \frac{2}{N} \sum_{\vk} 
 \frac{f(\lambda_{\vk}^{c2d}) - f(\lambda_{\vk + \vq}^{c2d})} {\lambda_{\vk}^{c2d} +\omega  - \lambda_{\vk + \vq}^{c2d}+ i \Gamma }  \,,
\label{koc2d}
\ee
and 
\be
\lambda_{\vk}^{c2d}= \xi_{\vk} - 2t_{z}  (\cos k_{x}a - \cos k_{y}a)^{2}  \cos k_{z}d \,.
\label{bandc2d}
\ee
Note that the $k_{z}$ summation in $\sum_{\vk}$ is changed to $-\pi \leq k_{z}d \leq \pi$. 

The LRC is also simplified for $c=2d$: 
\bea
&&\tilde{h}_{1}=0, \quad  \tilde{h}_{2}=\tilde{h}_{4}=4, \quad \tilde{h}_{3}=- 8  \, \\
&& V(\vq)= \frac{V_{c}}{{\rm det}\tilde{V}} \left[ 
\alpha ( 2 - \cos q_{x}a - \cos q_{y}a) +4 \right] \, \\
&& V^{'}(\vq)= \frac{V_{c}}{{\rm det}\tilde{V}} (4 \cos q_{z}d) \, \\
&& {\rm det} \tilde{V} = \left[ \alpha ( 2- \cos q_{x}a - \cos q_{y}a) + 4 (1- \cos q_{z}d) \right] 
\nonumber \\
&& \hspace{15mm} \times \left[ \alpha ( 2- \cos q_{x}a - \cos q_{y}a) + 4 (1+ \cos q_{z}d) \right]\,, 
\eea
and $V^{'}_{+}(\vq) = V^{'}(\vq)$ and $V^{'}_{-}(\vq) =0$ in \eq{Vpmp}. Therefore \eq{kqw2} is reduced to 
\bea
&&\kappa(\vq, \omega) = \frac{(\kappa^{0}_{11} +  \kappa^{0}_{\cos}) \left[ 
1- ( \kappa^{0}_{11} - \kappa^{0}_{\cos}) (V(\vq) - V^{'}(\vq)) \right] }
{\left[ 1- ( \kappa^{0}_{11} + \kappa^{0}_{\cos}) (V(\vq) + V^{'}(\vq))\right]  
\left[ 1- ( \kappa^{0}_{11} - \kappa^{0}_{\cos}) (V(\vq) - V^{'}(\vq))\right] } \,, \\
&& \hspace{14mm} = \frac{\kappa^{0}_{11} +  \kappa^{0}_{\cos}} {
1- ( \kappa^{0}_{11} + \kappa^{0}_{\cos}) (V(\vq) + V^{'}(\vq)) }\,.
\label{kqwc2d}
\eea
The interaction is calculated as 
\be
V(\vq) + V^{'}(\vq) = \frac{V_{c}/4} {\alpha ( 2 - \cos q_{x}a - \cos q_{y}a)/4 + 1- \cos q_{z}d} \,.
\label{Vc2d} 
\ee

On the other hand, one may start calculations by assuming $c=2d$ in \fig{bilayer} and take a smaller unit cell. In this case, the systems is reduced to a single-layer model per unit cell. Following the previous work \cite{becca96}, one can easily obtain the LRC as 
\be
V_{\rm single}(\vq) = \frac{V_{c}^{\rm single}}{
\alpha_{\rm single} (2- \cos q_{x}a - \cos q_{y}a) + ( 1 - \cos q_{z}d) } \,,
\label{V-single}
\ee
where $V_{c}^{\rm single} = \frac{e^{2}}{a^{2} (2 d)} \frac{d^{2}}{2 \epsilon_{\perp}}$ and $\alpha_{\rm single} = \frac{\epsilon_{\parallel}d^{2}}{\epsilon_{\perp}a^{2}}$. We here recall that we have defined $V(\vq)$ and  $V^{'}(\vq)$ by dividing the unit cell volume $a^{2}c$ in \eq{Vq-define}. Hence we also define  $V_{c}^{\rm single} $ by dividing the same volume as $a^{2}(2d)$ to make a direct comparison between different formalisms  possible.  

The charge response function is then given by 
\be
\kappa_{\rm single}(\vq, \omega) = \frac{\kappa^{0}_{\rm single}(\vq, \omega)} {
1- V_{\rm single}(\vq) \kappa^{0}_{\rm single}(\vq, \omega)} \,,
\label{kqw-single}
\ee
where 
\be
\kappa_{\rm single}^{0}(\vq, \omega) = \frac{2}{N} \sum_{\vk} \frac{ f(\xi_{\vk}^{\rm single}) - f(\xi_{\vk+\vq}^{\rm single})}{\xi_{\vk}^{\rm single} + \omega - \xi_{\vk+\vq}^{\rm single} + i \Gamma} \,,
\label{ko-single}
\ee
and 
\bea
&&\xi_{\vk}^{\rm single} = -2 t (\cos k_{x}a + \cos k_{y}a) -4t' \cos k_{x}a \cos k_{y}a 
-2t^{''} (\cos 2k_{x}a + \cos 2k_{y}a ) \nonumber \\ 
&& \hspace{16mm} -2t_{z}(\cos k_{x}a - \cos k_{y}a)^{2} \cos k_{z}d - \mu
\label{band-single}
\eea

Now we check that both approaches provide the same analytical results. The functional form of the dynamical charge susceptibility $\kappa(\vq, \omega)$ in \eq{kqwc2d} is the same as \eq{kqw-single}. We can check from Eqs.~(\ref{koc2d}), (\ref{bandc2d}), (\ref{ko-single}), and (\ref{band-single})
\be
\kappa^{0}_{11}+ \kappa^{0}_{\cos} = \kappa^{0}_{\rm single} \,,
\ee
with the same band dispersion. The interaction parts Eqs.~(\ref{Vc2d}) and (\ref{V-single}) are also identical because we can readily confirm $\alpha_{\rm single}= \alpha/4$ and $V_{c}^{\rm single}=V_{c}/4$.

\begin{figure}[t]
\centering
\includegraphics[width=16cm]{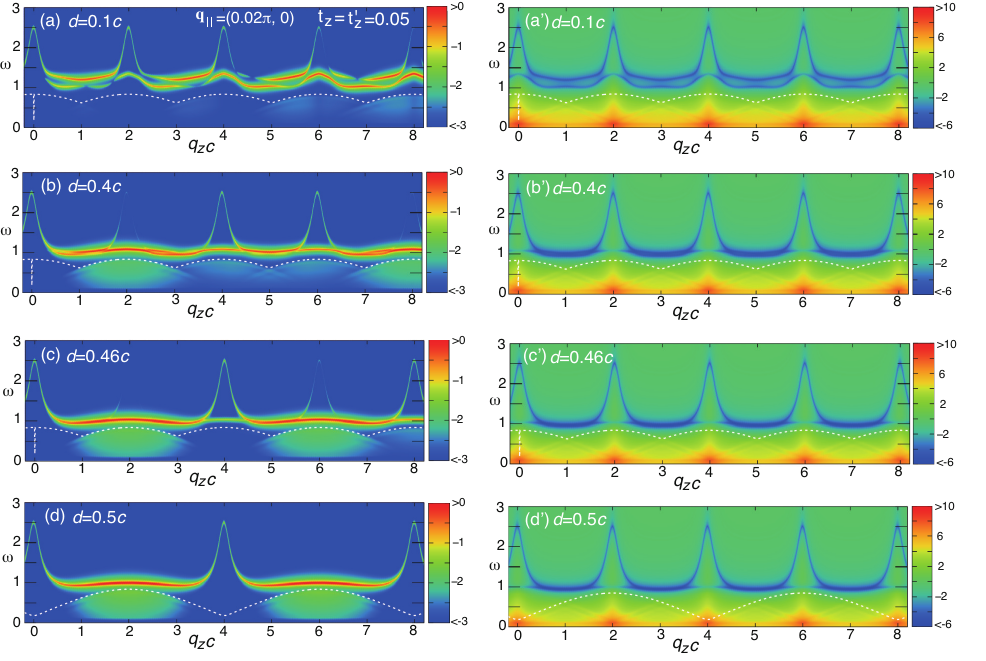}
\caption{Intensity map of $\log_{10} | {\rm Im}\kappa(\vq, \omega)|$ (left panels) and $\log_{10}|{\mathfrak{det}}|^{2}$ given in \eq{det2} (right panels) for various choices of intrabilayer distance $d$: (a) and (a') $d=0.1c$, (b) and (b') $d=0.4c$, (c) and (c') $d=0.46c$, and (d) and (d') $d=0.5c$. To get a reasonable contrast, the same color is used above 0 (10) and below -3 (-6) on the left (right) panels. The white dotted curve is the upper boundary of the continuum spectrum. The $\omega_{-}$ mode vanishes at $q_{z}c=0$ for $d\neq 0.5c$ and this special feature is reflected as a spike of the dotted curve there. 
}
\label{kqw-d}
\end{figure}

We have proved analytically that $\kappa(\vq, \omega)$ exhibits exactly $2 \pi$ periodicity with respect to $q_{z}d$ when $c=2d$ and $t_{z}=t_{z}^{'}$, namely, $4 \pi$ periodicity with respect to $q_{z}c$. It may be interesting to see how this periodicity changes as $d$ is closer to $c/2$. Fixing $t_{z}=t_{z}^{'}=0.05$ and $\vq_{\parallel}=(0.02\pi, 0)$, we compute Im$\kappa(\vq, \omega)$ as a function of $q_{z}c$ for a sequence of $d$ in \fig{kqw-d}. 

%%%%%%%%%%%%%%%%%%%%% FIG. 17 %%%%%%%%%%%%%%%%%%%%%%%%
%%%%%%%%%%%%%%%%%%%%%%%%%%%%%%%%%%%%%%%%%%%%%%%%%

In \fig{kqw-d}(a), both $\omega_{\pm}$ modes are clearly distinguished and we see approximately $2 \pi$ periodicity. The intensity of the $\omega_{+}$ mode becomes very weak near $q_{z}c=6\pi$ and $8\pi$ and apparently loses $2\pi$ periodicity. Nonetheless, we do not see a tendency that Im$\kappa(\vq, \omega)$ would exhibit $4 \pi$ periodicity. 
In \fig{kqw-d}(b), the difference between the $\omega_{\pm}$ modes is less visible as $d$ is closer to $0.5c$, 
especially away from $q_{z}c=2n\pi$. This is simply because we expect only one mode at $d=0.5c$, i.e., in a single-layer case. The intensity of the $\omega_{+}$ mode at $q_{z}c=2\pi$ and $8\pi$ is strongly suppressed. Yet we do not see a tendency toward $4\pi$ periodicity. At $d=0.46c$ shown in \fig{kqw-d}(c), we eventually start to find a tendency that Im$\kappa(\vq, \omega)$ could exhibit $4\pi$ periodicity and we finally obtain a perfect  $4\pi$ periodicity at $d=0.5c$ in \fig{kqw-d}(d). 

In the right panels in \fig{kqw-d} we plot the modulus of the denominator of $\kappa(\vq, \omega)$ [see Eqs.~(\ref{detk}) and (\ref{det2})]. While the minimum value becomes relatively large around $q_{z}c=2n\pi$ at $\omega \approx 1$ in Figs.~\ref{kqw-d}(b')(c'), it is due to a relatively large mixture of the continuum spectrum broadened by a finite $\Gamma$ [\eq{gmunu}]. We can check analytically that the denominator of $\kappa(\vq, \omega)$ [Eqs.~(\ref{detk}) and (\ref{kqw2})] has $2\pi$ periodicity with respect to $q_{z}c$, including the case of $d=0.5c$ as shown in \fig{kqw-d}(d'). A comparison with Figs.~\ref{kqw-d}(a)--(d) demonstrates that the numerator of $\kappa(\vq, \omega)$ determines the intensity of Im$\kappa(\vq, \omega)$ and eventually leads to the $4 \pi$ periodicity at $d=0.5c$.

\section{Split of \boldmath{$\omega_{\pm}$} modes at $q_{z}c=\pi$ for a small \boldmath{$t_{z}$}}

\begin{figure}[ht]
\centering
\includegraphics[width=8cm]{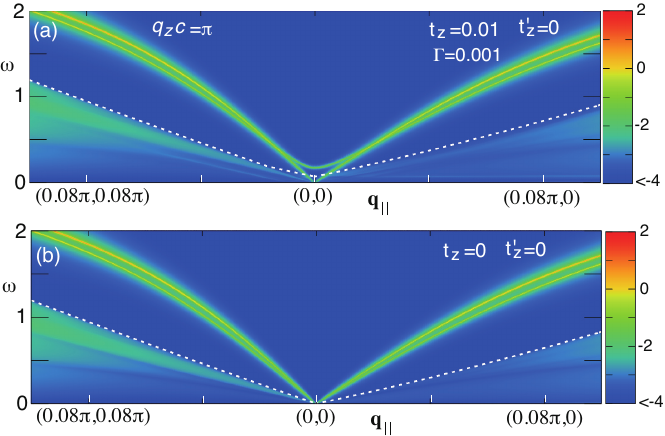}
\caption{Intensity map of $\log_{10} | {\rm Im}\kappa(\vq, \omega)|$ in the directions of $(0.1\pi, 0.1\pi)$-$(0,0)$-$(0.1\pi,0)$ for $q_{z}c=\pi$ at $t_{z}=0.01$ (a) and 0 (b); the broadening parameter  is taken as a smaller value $\Gamma=0.001$. The split of the $\omega_{\pm}$ modes is clearly visible. This figure should be compared with \fig{kqw-tz001}(c). The white dotted curve is the upper boundary of the continuum spectrum. To keep an appropriate contrast, the same color is used below -4. 
}
\label{kqw-qzpi-split}
\end{figure}

This appendix is supplementary to the discussion in the context of \fig{kqw-tz001}(c) and \fig{kqw-tz001detail}(b), where both $\omega_{\pm}$ modes seem nearly degenerate. However, there is a split between them, which may be less clear in the main text because we have taken a broadening parameter $\Gamma=0.01$. The split of the $\omega_{\pm}$ modes is more visible when we take a smaller $\Gamma=0.001$ as shown in \fig{kqw-qzpi-split}(a). This split comes from the difference of the effective interaction $V(\vq) \pm V_{+}(\vq)$ as explained in \eq{split-eq}, not from a small value of $t_{z}$. In fact, the same split is present also for $t_{z}=0$ as shown in \fig{kqw-qzpi-split}(b). In this case the $\omega_{+}$ mode also becomes gapless at $\vq_{\parallel}=(0,0)$. 
%%%%%%%%%%%%%%%%%%%%% FIG.  18 %%%%%%%%%%%%%%%%%%%%%%%%
%%%%%%%%%%%%%%%%%%%%%%%%%%%%%%%%%%%%%%%%%%%%%%%%%

\bibliography{main} 

\end{document}